\documentclass[showpacs,preprintnumbers,nofootinbib,
superscriptaddress,amsmath,floatfix,prd]{revtex4}

\usepackage{amssymb}  
\usepackage{graphicx}
\usepackage{exscale}
\usepackage{textcomp}
\usepackage{enumerate}
\usepackage{multirow} 

\usepackage[dvips]{color}

\usepackage[normalem]{ulem}  

\newcommand{\non}{\nonumber\\}

\newcommand{\be}{\begin{equation}}
\newcommand{\ee}{\end{equation}}
\newcommand{\bea}{\begin{eqnarray}}
\newcommand{\eea}{\end{eqnarray}}
\newcommand{\ba}[1]{\begin{array}{#1}}
\newcommand{\ea}{\end{array}}

\newcommand{\Tr}{{\rm Tr}}

\begin{document}

\title{Holographic baryonic matter in a background magnetic field}

\author{Florian Preis}
\email{fpreis@hep.itp.tuwien.ac.at}
\affiliation{Institut f\"{u}r Theoretische Physik, Technische Universit\"{a}t Wien, 1040 Vienna, Austria}

\author{Anton Rebhan}
\email{rebhana@hep.itp.tuwien.ac.at}
\affiliation{Institut f\"{u}r Theoretische Physik, Technische Universit\"{a}t Wien, 1040 Vienna, Austria}

\author{Andreas Schmitt}
\email{aschmitt@hep.itp.tuwien.ac.at}
\affiliation{Institut f\"{u}r Theoretische Physik, Technische Universit\"{a}t Wien, 1040 Vienna, Austria}

\begin{abstract}
We discuss the effect of baryonic matter on the zero-temperature chiral phase transition at finite chemical potential in the presence
of a background magnetic field.
The main part of our study is done in the deconfined geometry of the Sakai-Sugimoto model, i.e.\ at large $N_c$ and strong coupling, with non-antipodal 
separation of the flavor branes. 
We find that for not too large magnetic fields baryonic matter completely removes the chiral phase transition:
chirally broken matter persists up to arbitrarily large chemical potential.
At sufficiently large magnetic fields, baryonic matter becomes disfavored and mesonic matter is directly superseded by quark matter. 
In order to discuss the possible relevance of our results to QCD, we compute the baryon onset in a relativistic mean-field 
model including the anomalous magnetic moment and point out the differences to our holographic calculation.
\end{abstract}


\maketitle

\section{Introduction}
\label{sec:intro}

\subsection{Context}
\label{sec:context}

In the hadronic phase of Quantum Chromodynamics (QCD) the approximate $SU(N_f)_L\times SU(N_f)_R$ chiral symmetry is 
spontaneously broken by a chiral condensate. This condensate of quark-antiquark and quark-hole pairs is expected to melt 
for large temperatures $T$ and/or large baryon chemical potentials $\mu_B$. While the finite-temperature behavior at $\mu_B=0$ 
can be understood with the help of 
lattice calculations \cite{Aoki:2006we}, there are currently no first-principle methods to understand the fate of the chiral condensate
at small temperatures and large $\mu_B$. We do know from first principles that chiral symmetry is also spontaneously broken 
at {\it asymptotically} large 
$\mu_B$ -- albeit by a different mechanism, a diquark condensate in the color-flavor locked (CFL) phase \cite{Alford:1998mk,Alford:2007xm}. 
Studies of dense matter at large, but not asymptotically large, $\mu_B$ currently rely on model calculations. 
The logical possibilities regarding chiral symmetry, borne
out in these models, are either an intermediate chirally restored region, i.e., at least two phase transitions between ordinary nuclear 
matter and CFL matter \cite{Ruester:2005jc,Warringa:2006dk} or a continuous transition such that chiral symmetry is broken throughout 
the $T=0$ axis of the 
QCD phase diagram \cite{Schafer:1998ef,Hatsuda:2006ps,Schmitt:2010pf}. In this paper we shall ignore diquark condensation for simplicity and 
only consider chiral symmetry breaking via the ``usual'' mechanism.   

Phenomenologically this region of intermediate $\mu_B$ and small temperatures is relevant for the physics of neutron stars. The question of the 
chiral phase 
transition(s) in the QCD phase diagram is thus related to the question whether neutron stars are hybrid stars, i.e., whether they have a 
quark 
matter core surrounded by a mantle of nuclear matter, or whether they are entirely made out of nuclear (and possibly hyperonic) matter. 
Since neutron stars have large magnetic fields, it is interesting to investigate the chiral (and deconfinement) transitions in a large magnetic 
background. Many studies about the phases of QCD in a background magnetic field have recently been pursued, for 
instance on the lattice \cite{D'Elia:2010nq} as well as in model calculations \cite{Ebert:1999ht,Inagaki:2003yi,Boomsma:2009yk,Gatto:2010qs,Mizher:2010zb,Frolov:2010wn}.
From a simple estimate we can infer that extraordinarily large magnetic fields are required to have a sizable effect on QCD properties: if we 
simply convert the QCD scale into a scale for the magnetic field, we obtain 
$\Lambda_{\rm QCD}^2\sim (200\,{\rm MeV})^2\sim 2\times 10^{18}\, {\rm G}$.
This is large, but perhaps not out of reach for the cores of neutron stars whose surface magnetic fields have been measured to be as large as 
$\sim 10^{15}\, {\rm G}$ \cite{Duncan:1992hi}, possibly leading to magnetic fields in the core up to $\sim 10^{18}\, {\rm G}$ \cite{Lai} 
or, in the case of quark stars, even up to $\sim 10^{20}\, {\rm G}$ \cite{Ferrer:2010wz}. 
Besides this phenomenological motivation it is also of theoretical interest to consider large magnetic 
fields. For instance, it is instructive to consider the limit of asymptotically large magnetic fields which often becomes simple 
and which in turn may be used to understand fundamental properties of dense matter, also in a smaller magnetic background.       
  
\subsection{Purpose}

In this paper, we use the Sakai-Sugimoto model \cite{Sakai:2004cn,Sakai:2005yt} to study the chiral phase transition in a magnetic field. 
This model, making use of the gauge/gravity duality \cite{Maldacena:1997re,Gubser:1998bc,Witten:1998qj,Witten:1998zw}, is a top-down approach 
to large-$N_c$ QCD at low energies. In the original version of the model, $N_f$ right-handed and $N_f$ left-handed flavor branes are asymptotically
separated on antipodal points of the circle of a compactified extra dimension. In this version the phase structure is very simple, at least in the 
probe brane approximation $N_f\ll N_c$.
The non-antipodal, or ``decompactified'',  version, which corresponds to a (non-local) 
Nambu-Jona-Lasinio (NJL) model \cite{Antonyan:2006vw,Davis:2007ka}, is more susceptible to a chemical potential and a magnetic field, even 
in the probe brane approximation. Therefore, the phase structure is very rich and may even be more appropriate to learn something 
about real-world QCD with $N_c=3$. 

The present work is an extension of our previous study \cite{Preis:2010cq}. The main observation of ref.\ \cite{Preis:2010cq} was that, 
for certain intermediate values of the baryon chemical potential, chiral symmetry gets {\it restored} upon increasing the magnetic field. 
We have termed this effect ``Inverse Magnetic Catalysis'' (IMC) since it is in apparent contrast to ``Magnetic Catalysis'' (MC) 
\cite{Klimenko:1990rh,Gusynin:1994re} which, simply put, says that a magnetic field works in favor of a chiral condensate. The original 
works on MC
have pointed out that at weak coupling fermion-antifermion pairing in the presence of a magnetic field is not unlike Cooper 
pairing in a superconductor \cite{Gusynin:1995nb}, leading to a chiral condensate even for arbitrarily small attractive interaction. IMC 
at finite chemical potential can occur as follows. A nonzero 
chemical potential creates an asymmetry between fermions and antifermions and thus imposes a stress on the chiral condensate. Eventually, 
chiral symmetry gets restored at some critical chemical potential. The simple but important observation of 
ref.\ \cite{Preis:2010cq} was that the free energy 
cost to maintain the chiral condensate in spite of a nonzero chemical potential is proportional to the magnetic field. At weak coupling the gain 
in free energy from condensation grows faster with the magnetic field than this cost. As a consequence, at weak coupling a magnetic field can only break, 
never restore, chiral symmetry at a given chemical potential. This manifestation of MC is recovered in the Sakai-Sugimoto model 
for very large magnetic fields. For smaller magnetic fields, however, 
the opposite IMC 
is possible since the gain from condensation turns out to be less efficient in compensating the cost. This effect seems to be of 
general nature since it has also been observed for instance in NJL calculations 
\cite{Inagaki:2003yi,Boomsma:2009yk,Fayazbakhsh:2010bh,Chatterjee:2011ry}. Nevertheless, our previous study
was incomplete since we have for simplicity ignored baryonic matter. It {\it did} contain nonzero baryon number induced by a meson supercurrent via the axial anomaly \cite{Son:2007ny,Thompson:2008qw,Bergman:2008qv,Rebhan:2008ur} but not ``normal'' baryons in the following sense. 

Baryons are introduced in the Sakai-Sugimoto model as D4-branes wrapped on the $S^4$ of the background geometry \cite{Sakai:2004cn,Hata:2007mb},
implementing the general idea of baryons in the AdS/CFT context \cite{Witten:1998xy,Gross:1998gk}. Because of the flux carried by these D4-branes, 
each necessarily comes 
with $N_c$ string endpoints. The other end of each string is attached to the flavor D8-branes. In this way the AdS/CFT baryon realizes the usual
picture of a baryon as a colorless object made of $N_c$ quarks which carry color and flavor. 
In the non-supersymmetric Sakai-Sugimoto model, the strings pull the baryon D4-branes towards and ``into'' the D8-branes \cite{Callan:1999zf}. 
Thus it turns out that a baryon can equivalently be described as a gauge field configuration with nontrivial topological charge in the world-volume 
gauge theory on the D8-branes. 
The charge of the wrapped D4-brane is related to the topological charge carried by the instanton  \cite{Douglas:1995bn}. The latter, in turn, can be
interpreted as baryon charge since it couples to the $U(1)_B$ part of the gauge group in the bulk, associated with the global group of baryon 
number conservation in the corresponding field theory at the boundary. The interpretation 
of baryons as instantons yields a natural connection to the Skyrme model where baryons appear as solitons \cite{Skyrme:1961vq}. (This connection
is discussed in ref.\ \cite{Son:2003et} in the context of a five-dimensional gauge theory.) And indeed, it 
has been shown that the Sakai-Sugimoto model includes the Skyrme model in a certain limit \cite{Sakai:2004cn}. 

Spectra and properties of single baryons in the Sakai-Sugimoto model have been computed with a Yang-Mills action approximating the  
Dirac-Born-Infeld (DBI) action associated with the D8-branes
\cite{Sakai:2004cn,Hata:2007mb} (for analogous calculations in the non-antipodal version of the model see \cite{Seki:2008mu}).
This approximation
is not sufficient for our purpose since certain effects in a background magnetic field require the full DBI action,
for instance the phase transition in chirally restored quark matter which is reminiscent of a transition into the lowest Landau level
\cite{Preis:2010cq,Lifschytz:2009sz}. Using the DBI action, the description of baryons as instantons of a non-abelian gauge theory becomes 
extremely complicated. We thus restrict ourselves to single-flavor physics 
 where the gauge theory in the bulk becomes abelian. [Strictly speaking, at $N_f=1$ there is no spontaneous 
chiral symmetry breaking because the $U(1)_A$ 
is always broken due to the axial anomaly; however, at large $N_c$ this effect of the anomaly is suppressed.]
In this case,
baryons have been approximated as pointlike instantons or, equivalently, pointlike D4-branes in the antipodal and 
non-antipodal versions of the Sakai-Sugimoto model, see refs.\ \cite{Rozali:2007rx} and \cite{Bergman:2007wp}, respectively. In this 
sense one can consider the main part of the present study as an extension of the results of ref.\ \cite{Bergman:2007wp} to nonzero magnetic 
fields. Baryons in a magnetic field have been considered previously \cite{Bergman:2008qv}, however in the confined geometry with antipodal separation. Therefore, our study is also an extension of ref.\ \cite{Bergman:2008qv} to the non-antipodal deconfined case, where the phase structure is much richer, as we shall see. Our study has some overlap with refs.\ \cite{Burikham:2009yi,Burikham:2011ng,Burikham:2011rg}. While these references treat the baryonic 
phase purely numerically, we shall present analytical approximations which enable us to interpret the results physically. More importantly,
however, we disagree with several results in these works, see footnote \ref{burikham} in sec.\ \ref{sec:nonzero}.

\subsection{Holographic vs.\ real-world baryonic matter}

Using the holographic setup just described to determine the effect of baryonic matter on the chiral phase transition and 
on IMC requires us to ask whether we expect to be able to draw any conclusions for QCD. It is known that 
baryonic matter in the Sakai-Sugimoto model is different from real-world baryonic matter in at least one important aspect. Namely, 
holographic baryonic matter is not self-bound, i.e., not stable at zero pressure. This unrealistic property originates from the baryon-baryon
potential being repulsive for all distances \cite{Kaplunovsky:2010eh}. As a consequence, the zero-temperature onset of baryons is a second-order 
phase transition at the baryon mass $\mu_B=m_B$, and baryonic matter right after this onset has infinitesimally small density 
\cite{Rozali:2007rx,Bergman:2007wp}. In contrast, the onset of real-world baryonic matter is a first-order transition at $\mu_B=m_B+E_0$ with the 
binding energy $E_0\simeq -16\, {\rm MeV}$, and the density right after the onset is the nuclear ground state density 
$n_0\simeq 0.15 \,{\rm fm}^{-3}$.

In order to work out the differences between holographic vs.\ real-world baryonic matter in our context, we compute the baryon onset in the 
mean-field approximation of the Walecka model \cite{Walecka:1974qa} in the presence of a magnetic field for electrically charged and neutral 
baryons and for various values of the anomalous magnetic moment. The Walecka model is a relativistic, field-theoretical model usually employed to 
describe dense nuclear matter in the context of neutron stars. It is constructed 
to reproduce properties of nuclear matter at the ground state density $n_0$. 
(To describe nuclear matter at densities beyond $n_0$, a plethora of models exists, and it is an ongoing effort 
to exclude models for instance with the help of neutron star data.) In the simplest version of the Walecka model, baryon interactions are modelled
by the exchange of the scalar sigma  meson and the vector omega meson. Since the scalar meson exchange is responsible for the attractive 
interaction, the absence of a binding energy can be attributed to the
absence (heaviness) of the sigma. 
And indeed, in the Sakai-Sugimoto model the lightest scalar meson is, in contrast to the real world, much heavier than the lightest vector meson. 
It is not completely understood whether this property is a large-$N_c$ effect or whether the strong coupling limit and/or a model artifact also 
play a role. It seems that the large $N_c$ limit is at least partially responsible since 
generalizations of the Walecka model to large values of $N_c$ have shown that nuclear matter tends to become unbound essentially
as soon as $N_c$ is larger than three \cite{Bonanno:2011yr,Giacosa:2011uk}. 

Studies of nuclear matter in a magnetic field using Walecka-like models have been done previously in the context of neutron stars 
\cite{Broderick:2000pe,Broderick:2001qw,Rabhi:2011ej}. In these works, the conditions of beta-equilibrium and electric neutrality play a 
crucial role and thus the results are relatively complicated. Since we are interested in simpler, more fundamental questions we do not take into
account these conditions. As a consequence, our results reflect the ``pure'' effect of a magnetic field on a single baryon species with a 
given electric charge and anomalous magnetic moment. They are thus, although less realistic, also interesting on their own 
beyond being a comparison to the holographic model.

The remainder of the paper is organized as follows. In sec.\ \ref{sec:setup} we present the Sakai-Sugimoto framework and the solution of the 
equations of motion
for the baryonic phase. These solutions are then used in sec.\ \ref{sec:nuclear} to compute the transition between mesonic and baryonic phases
in the plane of magnetic field and chemical potential. In particular, we shall discuss the role of the meson supercurrent for this transition. 
Sec.\ \ref{sec:walecka} is devoted to the calculation of the onset of baryonic matter in a magnetic field within the Walecka model.
In the final section of the main part, sec.\ \ref{sec:phasediagram}, we come back to our holographic approach and 
present the zero-temperature phase diagram, containing the chiral phase transition in the presence of baryonic matter. Readers who are only 
interested in the main results should consult fig.\ \ref{figPD} and the brief summary of our observations in sec.\ \ref{sec:discussion2}.
Sec.\ \ref{sec:summary} summarizes all results and gives a brief outlook for future work.

\section{Setup}
\label{sec:setup}

In this section we discuss one-flavor hadronic matter in a background magnetic field in the deconfined, chirally broken phase 
of the Sakai-Sugimoto model. The ingredients for our starting 
point can all be found in previous works wherefore we shall be very 
brief in the derivation of the action and the equations of motion. Details about the mesonic part in the same notation as used here can be found in 
ref.\ \cite{Preis:2010cq} while for the baryonic part we closely follow ref.\ \cite{Bergman:2007wp}. For more general introductions into the 
Sakai-Sugimoto model see the original works \cite{Sakai:2004cn,Sakai:2005yt} or the review \cite{Peeters:2007ab} or applications of the model 
for instance in refs.\ \cite{Horigome:2006xu,Aharony:2006da,Aharony:2007uu}. 

\subsection{Action}

The action we consider in the following is
\be \label{action}
S = S'_0 + S_B \, . 
\ee
Here $S_B$ is the contribution from the baryon D4-branes wrapped on the $S^4$, and $S'_0$ is the gauge field action on the D8-branes without the 
baryonic terms. Since we consider one flavor there is one D8-brane and one $\overline{\rm D8}$-brane. They correspond to left- and 
right-handed fermions, respectively. In the chirally broken phase considered in this section, the two branes are connected and have a nontrivial 
embedding in the background geometry, which has to be determined dynamically. 
The prime denotes a modification to the Chern-Simons (CS) part of the action by certain boundary terms. These terms have  
been argued to be necessary in the presence of a magnetic field \cite{Bergman:2008qv} and have been discussed further in 
ref.\ \cite{Rebhan:2009vc}, where in particular their problems in the context of the chiral magnetic effect have been pointed out. Starting without 
this modification we have 
\be
S_0 = S_{\rm DBI} + S_{\rm CS}^{(0)} \, . 
\ee
The notation $S_{\rm CS}^{(0)}$ indicates that there is a second CS
contribution from the baryons which we shall call $S_{\rm CS}^{(1)}$, see eq.\ (\ref{SCS21}).
The DBI part is 
\be \label{DBI}
S_{\rm DBI} = T_8 V_4 \int d\tau\, d^3x \int_{U_c}^\infty dU\, e^{-\Phi}\sqrt{{\rm det}(g+2\pi\alpha'F)} \, . 
\ee
Here $g$ is the induced metric on the D8-branes in the deconfined geometry (see eq.\ (\ref{ds})), $F$ the field strength tensor of the $U(N_f)$ 
gauge theory on the branes, which in our case is abelian since we work at $N_f=1$. Furthermore, $\alpha'=\ell_s^2$ with the string length $\ell_s$, 
$T_8=(2\pi)^{-8}\ell_s^{-9}$ is the D8-brane tension, $V_4\equiv 8\pi^2/3$ the $S^4$ volume, 
and  $e^{\Phi} = g_s(U/R)^{3/4}$ the dilaton with the string coupling $g_s$ and the curvature radius $R$ of the background geometry. The 
background geometry is determined by $N_c$ D4-branes (not to be confused with the baryon D4-branes wrapping the $S^4$) and fixed throughout the paper since
we work in the probe brane approximation.  
The integration is over Euclidean four-dimensional space-time with imaginary time $\tau$ and over one half of the connected flavor branes 
parametrized by the holographic coordinate $U\in[U_c,\infty]$, where $U_c$ is the tip of the U-shaped, connected branes and $U=\infty$ 
is the holographic boundary where the gauge symmetry becomes global and corresponds to the chiral symmetry group. In the
presence of pointlike baryons, as discussed below, it turns out that there is a cusp in the profile of the branes at the tip $U=U_c$.
The CS part in the gauge $A_U=0$ is  
\be \label{SCS1}
S_{\rm CS}^{(0)} = \frac{N_c}{24\pi^2}\int d\tau\, d^3x\int_{U_c}^\infty dU\,A_\mu F_{U\nu}F_{\rho\sigma}\epsilon^{\mu\nu\rho\sigma} \, ,
\ee
where $\mu,\nu, \ldots = 0,1,2,3$ and $\epsilon_{0123}=+1$. 

Our ansatz for the gauge fields takes into account the nonzero boundary values 
$A_0(U=\infty)=\mu_q$ and $A_1(U=\infty)=-x_2 B$ where $\mu_q$ is the quark chemical potential and $B$ the modulus of the 
non-dynamical background magnetic field, pointing in the 3-direction. This ansatz requires a nonzero $A_3(U)$ to fulfil the equations of motion,
and even a nonzero boundary value $A_3(U=\infty)$ to minimize the free energy. This boundary value is interpreted as the gradient of a meson 
field which corresponds to a meson supercurrent \cite{Thompson:2008qw,Bergman:2008qv,Rebhan:2008ur}. 
Within this ansatz and including the modification of the CS action mentioned above, we exactly 
follow the steps in ref.\ \cite{Preis:2010cq} to obtain
\be \label{S0prime}
S_0' = \frac{\cal N}{2}\frac{V}{T} \int_{u_c}^{\infty} du\,{\cal L}_0 \, , 
\ee
where $V$ is the three-volume, $T$ the temperature, and we have abbreviated
\be
{\cal N}\equiv \frac{N_c}{6\pi^2}\frac{R^2}{(2\pi\alpha')^3}  \, , 
\ee
and  
\be \label{L0}
{\cal L}_0 \equiv \sqrt{(u^5+b^2u^2)(1+fa_3'^2-a_0'^2+u^3fx_4'^2)} +
\frac{3}{2}\,  b  (a_3a_0'-a_0a_3') \, ,
\ee
with the first and second term coming from the DBI and CS contribution, respectively.
We have introduced the following dimensionless quantities: the gauge fields $a_\mu\equiv \frac{2\pi\alpha'}{R} A_\mu$, the magnetic field
$b\equiv 2\pi\alpha' B$, the holographic coordinate $u\equiv \frac{U}{R}$, and the coordinate of the 
compactified extra dimension $x_4\equiv \frac{X_4}{R}$, $X_4\in[0,\frac{2\pi}{M_{\rm KK}}]$ with the Kaluza-Klein mass $M_{\rm KK}$.
The D8- and $\overline{\rm D8}$-branes are asymptotically, i.e., for $u\to\infty$, separated in the $x_4$ direction by a distance $L$ which 
we shall take to be small with respect to the maximal separation on antipodal points of the circle, $L\ll \frac{\pi}{M_{\rm KK}}$.  
Moreover, the prime denotes derivative with respect to $u$, and 
\be \label{defuT} 
f(u) = 1-\frac{u_T^3}{u^3} \, , \qquad u_T^{1/2} = \frac{4\pi}{3} TR \, .
\ee
To obtain eq.\ (\ref{S0prime}) we have used the relation $R^3=\pi g_s N_c\ell_s^3$ and the induced metric on the D8-branes
\bea \label{ds}
ds^2 &=&  u^{3/2}[f(u)d\tau^2+\delta_{ij}dx^idx^j] +R^2\left\{\left[\frac{1}{u^{3/2}f(u)}+u^{3/2} x_4'^2\right]du^2+u^{1/2}d\Omega_4^2\right\} \, , 
\eea
where $d\Omega_4^2$ is the metric of the $S^4$.

Now we turn to the baryonic part,
\be \label{SB}
S_B = S_{\rm D4} + S_{\rm CS}^{(1)} \, .
\ee
To understand the origin of $S_{\rm CS}^{(1)}$ we must take a detour via the nonabelian case $N_f>1$. 
In that case, eq.\ (\ref{SCS1}) is not the only CS contribution. The integrand of eq.\ (\ref{SCS1}) is of the form $A\,F^2$ 
with a gauge field $A$ and field strength $F$ from 
the $U(1)$ part of the bulk gauge group $U(N_f)\cong SU(N_f)\times U(1)$. In general, there is also a contribution where the integrand has the 
form $A\,\Tr \,F^2$ where $A$ is from the abelian part, but $F$ from the nonabelian part with the trace taken over flavor space 
(see for instance Refs.\ \cite{Rebhan:2008ur,Hata:2007mb} for an explicit form of the CS action in terms of abelian and nonabelian parts). 
Baryon number is generated from the term which couples to $A_0$, i.e., we are interested in a contribution of the form  
\be  \label{SCS21}
S_{\rm CS}^{(1)} = \frac{N_c}{8\pi^2}\int_{\mathbb{R}^4\times {\cal U}}  A_0\, {\rm Tr}\,F^2 \, .
\ee
Here the integration goes over Euclidean space-time $\mathbb{R}^4$ and the holographic direction ${\cal U}$, parametrized by 
$(\tau, x^1,x^2,x^3,u)$. From this expression we see that the instanton charge,  
\be
N_4 = - \frac{1}{8\pi^2}\int_{\mathbb{R}^3\times {\cal U}} {\rm Tr}\,F^2 \, ,
\ee
with integration over three-dimensional space and holographic direction, parametrized by $(x^1,x^2,x^3,u)$, 
is identical to the baryon charge. For $N_f=1$, there is no non-singular instanton solution, and we use a pointlike approximation 
localized at $u=u_c$ and homogeneous in three-space \cite{Bergman:2007wp} (for attempts towards inhomogeneous solutions see 
for instance refs.\ \cite{Kim:2007vd,Rho:2009ym}),   
\be
\frac{1}{8\pi^2}\Tr \, F^2 = - \frac{N_4}{V}\delta(u-u_c) d^3x\,du \, . 
\ee
Thus the CS contribution becomes 
\be \label{SCS2}
S_{\rm CS}^{(1)} = -\frac{V}{T} {\cal N}\,n_4a_0(u_c)\, ,
\ee
where we have defined the dimensionless baryon charge density
\be \label{n4def}
n_4\equiv \frac{N_c}{\cal N}\frac{N_4}{V}\frac{R}{2\pi\alpha'} \, .
\ee

The first term on the right-hand side of eq.\ (\ref{SB}) is the action of $|N_4|$ D4-branes wrapped on the $S^4$, corresponding to 
$|N_4|$ (anti-)baryons. In the case of anti-baryons, these are $\overline{\rm D4}$-branes with negative charge $N_4<0$. 
In our approximation of pointlike D4-branes,
\be \label{SD4}
S_{\rm D4} = |N_4|T_4\int d\Omega_4d\tau\,e^{-\Phi}\sqrt{{\rm det} \,g} = \frac{V}{T}{\cal N} \frac{u_c}{3}\sqrt{f(u_c)} \,|n_4|\, , 
\ee
where $T_4=(2\pi)^{-4}\ell_s^{-5}$ is the D4-brane tension and where, in the second step, we have used the explicit 
forms of the metric and the dilaton.  For later use it is convenient to write $S_{\rm D4}$ in an alternative way, introducing the 
(dimensionful) mass $M_q$ of a constituent quark within the baryon,
\be \label{Mq}
S_{\rm D4} = \frac{V}{T}\frac{N_c|N_4|}{V} M_q \, , \qquad M_q \equiv \frac{R}{2\pi\alpha'}\frac{u_c}{3}\sqrt{f(u_c)} \, , 
\ee
such that $N_cM_q$ is the baryon mass.

We can now put together the contributions (\ref{S0prime}), (\ref{SCS2}), and (\ref{SD4}) to obtain the total action
\be
S = {\cal N}\frac{V}{T}\int_{u_c}^\infty du\, {\cal L}_0 + {\cal N}\frac{V}{T}\, n_4\left[{\rm sgn} (n_4) \frac{u_c}{3}\sqrt{f(u_c)}
-a_0(u_c)\right] \, , 
\ee
with ${\cal L}_0$ given in eq.\ (\ref{L0}). 

\subsection{Equations of motion}
\label{sec:eom}

From now on we shall set $f(u)\simeq 1$. This simplifies the analysis of the chirally broken phase tremendously since only within this approximation 
we can find semi-analytical solutions. As a consequence, our results are more transparent compared to a purely
numerical analysis and the physical interpretation becomes simpler. This approximation can be read in two ways because $f(u)\simeq 1$ is 
equivalent to $u_T\ll u_c$ (since $u\ge u_c$), and $u_T$ and $u_c$ are determined by independent scales of the model. From eq.\ (\ref{defuT})
we see that $u_T$ is proportional to the temperature. Hence $f(u)\simeq 1$ can be interpreted as the zero-temperature limit. We need to 
remember, however, that in the Sakai-Sugimoto model the confined phase is the ground state for $T<T_c=\frac{M_{\rm KK}}{2\pi}$. Therefore, we
can only take the $T\to 0$ limit of the deconfined phase if simultaneously we also let $M_{\rm KK}\to 0$, i.e., let the radius 
of the compactified extra dimension go to infinity.  The second option to fulfil $u_T\ll u_c$ is to change $u_c$ at fixed $u_T$. 
This can be achieved by decreasing the asymptotic separation of the D8- and $\overline{\rm D8}$-branes, i.e., by letting $L\to 0$. 
Geometrically, however, this is not
so different from the first limit: increasing the radius of the circle along which the branes are separated while keeping their distance fixed 
corresponds to decreasing the separation while keeping the radius fixed.
In either case we 
arrive at a ``decompactified'' limit where the gluon dynamics is decoupled 
from the system and which is not unlike a nonlocal NJL model on the 
field theory side \cite{Antonyan:2006vw,Davis:2007ka}.

The following parts, including the calculation in appendix \ref{app1}, and the next subsection \ref{sec:free}, are the most technical ones 
of this paper. They serve to collect all necessary results which shall be evaluated and discussed in secs.\ \ref{sec:nuclear} and 
\ref{sec:phasediagram}.

With $f(u)\simeq 1$ the equations of motion become in the integrated form 
\begin{subequations} \label{EOM}
\bea
\frac{a_0'\sqrt{u^5+b^2u^2}}{\sqrt{1+a_3'^2-a_0'^2+u^3x_4'^2}} &=& 3ba_3 + n_4 \, ,\label{EOMa0}\\[1.5ex]
\frac{a_3'\sqrt{u^5+b^2u^2}}{\sqrt{1+a_3'^2-a_0'^2+u^3x_4'^2}} &=& 3ba_0 + d \, ,\label{EOMa3}\\[1.5ex]
\frac{u^3\,x_4'\sqrt{u^5+b^2u^2}}{\sqrt{1+a_3'^2-a_0'^2+u^3x_4'^2}} &=& k \label{EOMx4}\, .
\eea
\end{subequations}
Here, $d$ and $k$ are integration constants, to be determined. The integration constant of the first equation is fixed to be $n_4$
because of the pointlike charge sitting at $u=u_c$. As written, these equations are identical to the 
ones without baryonic matter. The singularity at the tip of the brane, however, gives rise to different boundary conditions. 

Without explicitly solving the equations of motion we may compute the nonzero components of the four-current 
${\cal J}^\mu$. The current is given by  
\be
{\cal J}^\mu = -\left.\frac{\partial {\cal L}_0}{\partial A_\mu'}\right|_{u=\infty} \, , 
\ee
since the baryonic part $S_B$ does not contribute. 
(Here we have used the dimensionful gauge field $A_\mu$ to obtain a dimensionful current.) This yields
\bea \label{current}
{\cal J}^0 &=& \frac{2\pi\alpha'}{R} {\cal N}
\left(\frac{3}{2}b\jmath + n_4 \right) \, , \qquad {\cal J}^3 = -\frac{2\pi\alpha'}{R} {\cal N} \left(\frac{3}{2}b\mu  + d\right) \, ,
\eea
where we have used the equations of motion and where $\mu=a_0(\infty)$ is the dimensionless quark chemical potential, 
$\mu = \frac{2\pi\alpha'}{R} \mu_q$, and $\jmath = a_3(\infty)$ is the dimensionless meson supercurrent which has to be determined dynamically from 
minimization of the free energy. The dimensionless baryon density, 
\be \label{n1}
n = \frac{3}{2}b\jmath + n_4  \, ,
\ee
thus receives contributions from the meson supercurrent and from ``normal'' baryons\footnote{Note that the factor $\frac{2\pi\alpha'}{R}{\cal N}$, which turns $n$ into a dimensionful density, scales with $N_c$ (while $n$ does not). Therefore, although it may seem a bit confusing, $n$ is a (dimensionless) {\it baryon} density, while ${\cal J}^0$ is the (dimensionful) {\it quark} density. (In our previous work \cite{Preis:2010cq} the terminology was more sloppy since we referred to $n$ as a dimensionless quark density.)}. In our model, both 
contributions are of topological nature since they both originate from the CS term. We shall see below that $n$ is indeed the 
baryon number density also in the thermodynamical sense, obtained from taking the derivative of the free energy with respect to $\mu$, see 
eq.\ (\ref{check}). This consistency is, in the presence of a magnetic field, only achieved by the modification $S_0\to S_0'$ introduced in ref.~\cite{Bergman:2008qv}.
Analogously, ${\cal J}^3$ can also be obtained by taking the derivative of the free energy with respect to the corresponding source, which 
in this case is the supercurrent $\jmath$. Consequently, the condition that $\jmath$ minimize the free energy is equivalent to ${\cal J}^3=0$.
Nevertheless, one should not think of ${\cal J}^3=0$ as an externally imposed constraint; the system rather finds 
its minimal free energy for any given $\mu$ and $b$, and ${\cal J}^3=0$ is a mere consequence. 
Using ${\cal J}^3=0$ in Eq.\ (\ref{current}), we can immediately read off
\be \label{d}
d = -\frac{3}{2} b\mu \, .
\ee
We also need to minimize the on-shell action with respect to the parameters $n_4$ and $u_c$.
By using partial integration and the equations of motion, we have
\be
0 = \frac{\partial S_{\rm on-shell}}{\partial n_4} 
=\left(\sum_{i=0,3}\frac{\partial {\cal L}_0}{\partial a_i'}\frac{\partial a_i}{\partial n_4}
+\frac{\partial {\cal L}_0}{\partial x_4'}\frac{\partial x_4}{\partial n_4}\right)_{u=u_c}^{u=\infty}  \pm \frac{u_c}{3}-a_0(u_c)
-n_4\frac{\partial a_0(u_c)}{\partial n_4}  \label{Dn4} \, ,
\ee
where the upper (lower) sign holds for $n_4>0$ ($n_4<0$). The derivatives involving $x_4$ 
vanish since we keep the asymptotic separation of the flavor branes fixed, $x_4(\infty)-x_4(u_c)=\frac{\ell}{2}$ with 
the dimensionless separation $\ell=\frac{L}{R}$. The derivatives involving the 
gauge fields vanish at $u=\infty$ since $a_0(\infty)=\mu$, $a_3(\infty)=\jmath$ are held fixed. We also impose $a_3(u_c)=0$ such that the 
only nonvanishing term from the parentheses comes from the derivative involving $a_0$, at $u=u_c$. The equation then reduces to the simple condition
\bea \label{auc}
a_0(u_c)= \pm\frac{u_c}{3}  \, . 
\eea
The minimization with respect to $u_c$ becomes 
\bea
0 &=& \frac{\partial S_{\rm on-shell}}{\partial u_c}= 
\left(\sum_{i=0,3}\frac{\partial {\cal L}_0}{\partial a_i'}\frac{\partial a_i}{\partial u_c}
+\frac{\partial {\cal L}_0}{\partial x_4'}\frac{\partial x_4}{\partial u_c}\right)_{u=u_c}^{u=\infty}-{\cal L}_0(u=u_c)  
 \, , \label{duc}
\eea
where we have used eq.\ (\ref{auc}). Here we have to be more careful at the lower boundary because at this 
boundary $a_0$, $a_3$ and $x_4$ depend on $u_c$ explicitly as well as through the variable $u$. Therefore, 
\be
\frac{\partial a_0(u_c)}{\partial u_c}= a_0'(u_c) + \left.\frac{\partial a_0(u)}{\partial u_c}\right|_{u=u_c} \, ,
\ee
and the same for $a_3$ and $x_4$. Here, the left-hand side denotes the total derivative with respect to $u_c$. In the first term on the 
right-hand side the derivative acts on the dependence through $u$, while in the second term it acts on the explicit dependence. 
Making use of this relation and of the equations of motion, eq.\ (\ref{duc}) can after some algebra be written as
\bea \label{force}
\frac{|n_4|}{3} = \frac{\sqrt{g(u_c)}}{u_c^{3/2}} \, ,
\eea
where we have abbreviated 
\be\label{g}
g(u) \equiv u^8+b^2u^5-k^2-u^3[(bu_c\pm d)^2-n_4^2]\, .
\ee
The condition (\ref{force}) can be interpreted as a force balance equation which says that the force from the D8-brane tension cancels the 
one from the weight of the baryon D4-branes \cite{Bergman:2007wp}. This equation is now used to determine $k$, 
\be \label{k21}
k^2 = u_c^8+b^2u_c^5-u_c^3[(bu_c\pm d)^2-n_4^2]-u_c^3\left(\frac{n_4}{3}\right)^2 \, . 
\ee
In appendix \ref{app1} we discuss the explicit solution of the equations of motion in detail. Here we only
summarize the results. The gauge fields become
\begin{subequations} \label{solutions}
\bea
a_0(y) &=& \frac{\mu}{2}\left(1+\frac{\sinh y}{\sinh y_\infty}\right)+\left(\frac{\mu}{2}\mp \frac{u_c}{3}\right)\cosh y_\infty
\left(\frac{\sinh y}{\sinh y_\infty}-\frac{\cosh y}{\cosh y_\infty}\right) \, , \hspace{1cm}\\[1.5ex]
a_3(y) &=& \frac{\mu}{2}\frac{\cosh y-1}{\sinh y_\infty}+\left(\frac{\mu}{2}\mp \frac{u_c}{3}\right)\cosh y_\infty
\left(\frac{\cosh y-1}{\sinh y_\infty}-\frac{\sinh y}{\cosh y_\infty}\right) \, ,
\eea
\end{subequations}
with the new variable 
\be
y(u) = 3b\int_{u_c}^u\frac{v\,dv}{\sqrt{g(v)}} \, , 
\ee
and $y_\infty\equiv y(u=\infty)$. The embedding of the flavor branes is  
\be 
x_4(u) = k\int_{u_c}^u\frac{dv}{v^{3/2}\sqrt{g(v)}} \, .  
\ee
With $d$ from eq.\ (\ref{d}) and 
\be \label{n4}
n_4 = \frac{3b}{2\sinh y_\infty}\left[\mu + \left(\mu\mp \frac{2u_c}{3}\right)\cosh y_\infty\right] \, , 
\ee
$k$ and thus $g(u)$ can be expressed in terms of $u_c$ and $y_\infty$ which have to be determined numerically from the two coupled equations
\be \label{yinfell}
y_\infty = 3b\int_{u_c}^\infty \frac{ u^{3/2}du}{\sqrt{g(u)}} \, , \qquad \frac{\ell}{2} = k\int_{u_c}^\infty \frac{du}{u^{3/2}\sqrt{g(u)}} \, .
\ee
Since $y_\infty$ is manifestly positive, eq.\ (\ref{n4}) shows that $\mu$ and $n_4$
must have the same sign, as it should be. Note also that in $k$ and $g(u)$ the sign of $n_4$ appears only in a product with $\mu$. 
Therefore, the equations for $y_\infty$ and $u_c$ only depend on $|\mu|$. We may thus 
restrict ourselves in the following to positive $n_4$ 
without loss of generality.

Finally, the supercurrent becomes
\be \label{j}
\jmath = \frac{u_c}{3}\frac{\cosh y_\infty -1}{\sinh y_\infty} \, ,
\ee
while the total density is
\be \label{n}
n = 3b\left(\mu-\frac{u_c}{3}\right)\frac{1+\cosh y_\infty}{2\sinh y_\infty} \, .
\ee



\subsection{Free energy}
\label{sec:free}

The free energy of the baryonic phase is 
\be
\Omega_\vee = \frac{T}{V} S_{\rm on-shell} = {\cal N}\int_{u_c}^\infty du\, {\cal L}_0 \, , 
\ee
since the baryonic part $S_B$ does not contribute, due to eq.\ (\ref{auc}). (Of course, the effect of the baryonic part is implicitly 
present through the equations of motion.) By inserting the explicit solution -- most conveniently in the form of eqs.\ (\ref{a0a3}) -- we find
\bea
\Omega_\vee &=& {\cal N}\int_{u_c}^\infty du\, u^{3/2}\frac{u^5+b^2u^2-\frac{1}{2}[(bu_c+d)^2-n_4^2]}{\sqrt{g(u)}} 
+ {\cal N}\left(\frac{n_4 u_c}{6}- \frac{\mu n}{2}\right) 
 \, .
\eea
This form of the free energy is useful for a comparison with the purely mesonic phase, see eq.\ (3.14) in ref.\ \cite{Preis:2010cq}. 
For a proof that the derivative of $\Omega_\vee$ with respect to $\mu$ is indeed the (negative of the) total density we rewrite
\be \label{Orewrite}
\frac{\Omega_\vee}{\cal N} = \int_{u_c}^\infty du\,\frac{\sqrt{g(u)}}{u^{3/2}}+\frac{k\ell}{2}+\frac{y_\infty}{6b}[(bu_c+d)^2-n_4^2]- \frac{\mu n}{2} 
+\frac{n_4 u_c}{6} \, .
\ee
Now we note that $\Omega_\vee = \Omega_\vee[\mu,b,u_c(\mu,b),y_\infty(\mu,b)]$ and take the derivative with respect to $\mu$ at fixed $b$,
including the implicit dependence through $u_c$ and $y_\infty$. We obtain
\be
\frac{1}{\cal N}\frac{\partial\Omega_\vee}{\partial\mu} = -\frac{n_4}{3}\frac{\partial u_c}{\partial\mu}+\frac{(bu_c+d)^2-n_4^2}{6b}
\frac{\partial y_\infty}{\partial\mu} + \frac{\partial}{\partial\mu}\left(- \frac{\mu n}{2} 
+\frac{n_4 u_c}{6}\right) \, .
\ee
With the explicit expressions for $n_4$ and $n$ in eqs.\ (\ref{n4}) and (\ref{n}) one computes
\be
\frac{\partial}{\partial\mu}\left(- \frac{\mu n}{2} +\frac{n_4 u_c}{6}\right) = -n +\frac{n_4}{3}\frac{\partial u_c}{\partial\mu}
-\frac{(bu_c+d)^2-n_4^2}{6b}\frac{\partial y_\infty}{\partial\mu} \, , 
\ee
such that indeed
\be \label{check}
-\frac{\partial\Omega_\vee}{\partial\mu_q} = \frac{2\pi\alpha'}{R}{\cal N}n = \, {\cal J}^0 \, . 
\ee

\section{Holographic baryon onset}
\label{sec:nuclear}

\begin{figure} [t]
\begin{center}
\hbox{\includegraphics[width=0.45\textwidth]{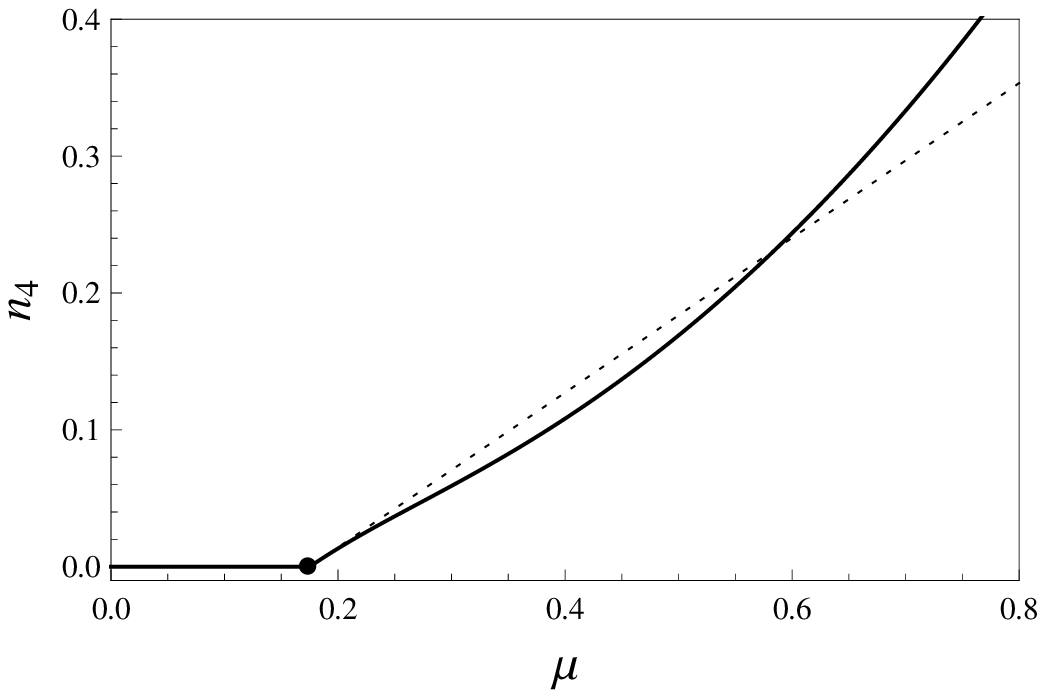}\hspace{0.3cm}
\includegraphics[width=0.45\textwidth]{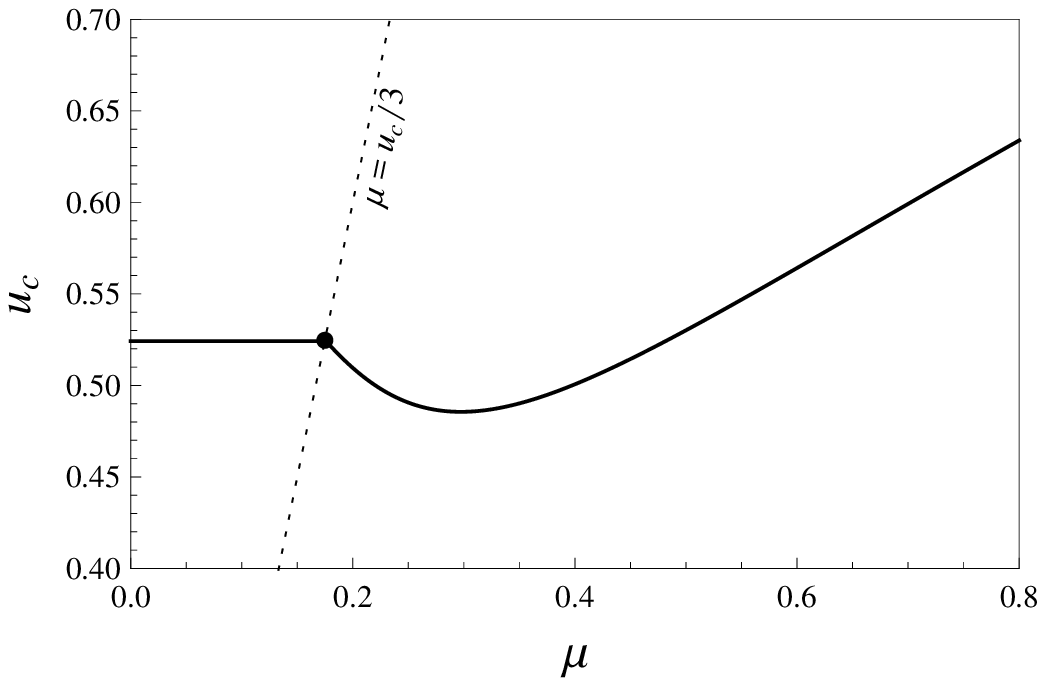}}
\caption{Solid lines: baryon density $n_4$ (left) and location of the tip of the connected flavor branes $u_c$ (right) at vanishing magnetic field. 
The baryonic phase becomes favored for 
$\mu> u_c/3$, indicated by the dotted line in the right panel. 
For chemical potentials smaller than that value, the mesonic 
phase is the ground state, in which -- at vanishing magnetic field -- the baryon density vanishes and $u_c$ is constant.
The dotted line in the left panel is the linear approximation (\ref{n4onset}). Here and in all other plots
we have set $\ell=1$ for notational convenience. The dependence on $\ell$ is always simple and corresponds to a rescaling of the axes, in 
this plot $\mu\to \ell^2 \mu$, $u_c\to \ell^2u_c$, $n_4\to \ell^5 n_4$.}
\label{figb0}
\end{center}
\end{figure}

\subsection{Zero magnetic field}
\label{sec:zeroB}

Let us start with the simplest case of a vanishing magnetic field. The numerical solution of eqs.\ (\ref{yinfell}) for this case yields the results
shown in fig.\ \ref{figb0}. 
For chemical potentials smaller than $\frac{u_c}{3}$ no solution with $n_4\neq 0$ exists, and the mesonic phase is the ground state. 
This phase has already been discussed in ref.\ \cite{Preis:2010cq} and is obtained by omitting the baryonic
contribution $S_B$ in the action (\ref{action}). To find the transition between the mesonic and baryonic phases, the ``baryon onset'', we consider 
the respective solutions wherever they exist and compare the respective free energies. (The mesonic and baryonic phases are not distinguished
by any symmetry, the baryon onset is only a phase transition in the sense that the second derivative of the free energy is discontinuous.) 
The numerical calculation shows that
as soon as a baryonic phase with $n_4>0$ exists it has lower free energy than the mesonic phase. This statement holds for all 
magnetic fields $b$.  

Before we come to the onset at finite $b$ in detail, we discuss the results for the $b=0$ case analytically. From the numerical solution
we read off that $y_\infty$ is linear in $b$ for small $b$. Thus we insert the ansatz $y_\infty = 3by_\infty^{0}$, with $y_\infty^{0}$
to be determined, into eqs.\ (\ref{n4}) and (\ref{n}) to find
\be \label{n4approx}
n_4(b=0)= n(b=0) = \frac{\mu-m_q^0}{y_\infty^{0}} \, ,
\ee
where $m_q^0 = \frac{2\pi\alpha'}{R}M_q^0$ is the dimensionless constituent quark mass at vanishing magnetic field,
$m_q^0 = \frac{u_c^0}{3}$ with $u_c^{0}=u_c(b=0)$. Note that $y_\infty^{0}$, $u_c^{0}$ and thus also $m_q^0$ are still complicated functions of 
$\mu$, to be computed numerically in general. From eq.\ (\ref{n4approx}) we conclude that a baryon chemical potential at least as large as the 
baryon mass is needed to obtain a nonzero baryon density. This shows that there is no binding energy for our holographic baryonic matter, since then 
the onset chemical potential would be lower than the mass, as it is the case for real-world nuclear matter. As a consequence, our baryon onset is a 
second-order, not first-order, phase transition. We shall see that this is also true for nonvanishing magnetic fields. 

In the vicinity of the onset we may compute the density analytically. First, by inserting $n_4=0$ into eqs.\ (\ref{yinfell}), we obtain at the 
onset,  
\be \label{uc0yinf0}
u_{c,{\rm onset}}^{0} = \frac{16\pi}{\ell^2}\left[\frac{\Gamma\left(\frac{9}{16}\right)}{\Gamma\left(\frac{1}{16}\right)}\right]^2 \, , \qquad
y_{\infty,{\rm onset}}^{0} = \frac{\ell^3}{512\pi} \frac{\Gamma\left(\frac{3}{16}\right)}{\Gamma\left(\frac{11}{16}\right)}
\left[\frac{\Gamma\left(\frac{1}{16}\right)}{\Gamma\left(\frac{9}{16}\right)}\right]^3 \, , 
\ee
and thus 
\be
\mu_{\rm onset}^{0} = m_q^0 \simeq \frac{0.17}{\ell^2} \, .
\ee
This result is in agreement with ref.\ \cite{Bergman:2007wp}, see for instance fig.\ 10 in this reference. 
For chemical potentials larger than but close to $\mu_{\rm onset}^{0}$, the dimensionless density behaves as (see appendix \ref{app2a} for the 
derivation)
\be \label{n4onset}
n_4(b=0) =  \frac{2\ell(m_q^0)^2}{\xi} \,(\mu-m_q^0) + {\cal O}[(\mu-m_q^0)^2] \, ,
\ee
where
\be \label{xi}
\xi\equiv \frac{1}{9}\left[\pi\frac{\Gamma\left(\frac{1}{16}\right)\Gamma\left(\frac{3}{16}\right)}
{\Gamma\left(\frac{9}{16}\right)\Gamma\left(\frac{11}{16}\right)}
 -\frac{2}{9}\right] \simeq 0.11 \, .
\ee
We plot the linear approximation (\ref{n4onset}) in comparison to the full result in the left panel of fig.\ \ref{figb0}. 

This linear behavior might seem unexpected because for fermions in 3+1 dimensions, at least in the non-interacting case, one 
would have $n_4\sim(\mu-m)^{3/2}$ just above the onset. It is therefore interesting to compare $n_4$ from  eq.\ (\ref{n4onset}) 
with the density of a Bose condensate instead. To this end, consider a  
$\phi^4$ model with chemical potential at zero temperature, 
$\Omega = \frac{m^2-\mu^2}{2}\phi^2 +\frac{\lambda}{4}\phi^4$. Minimizing $\Omega$ with respect to the 
condensate $\phi$ and inserting the result into the density $n=-\frac{\partial\Omega}{\partial\mu}$ yields $n=\mu\frac{\mu^2-m^2}{\lambda}=
\frac{2m^2}{\lambda}(\mu-m) + {\cal O}[(\mu-m)^2]$. The linear term looks exactly like the one in eq.\ (\ref{n4onset}) 
(in the $\phi^4$ model, the quadratic term is positive; this is in contrast to our holographic result, as fig.\ \ref{figb0} shows). 
The similarity of the baryon density with the density of a Bose condensate is not too surprising because we work in the large-$N_c$ limit.
In this limit, changing the number of constituent quarks from even to odd, i.e., from $N_c$ to $N_c+1$ does not make a difference. Therefore, 
in our context one may very well talk about a condensate of baryons. This terminology is for instance used in ref.\ \cite{Rozali:2007rx}.
It is instructive to express eq.\ (\ref{n4onset}) in terms of dimensionful quantities, see first row of table~\ref{densitytable},
where we show the dimensionful quark density ${\cal J}^0$. In the case $B=0$ this density only receives a contribution from ``normal'' baryons,
${\cal J}^0(B=0) = N_cN_4/V$, see eqs.\ \eqref{n4def} and (\ref{current}). Comparing this expression with the $\phi^4$ model, we may identify
$\frac{3\xi}{2}\lambda\frac{\pi/M_{\rm KK}}{L}$ with a coupling constant for an effective repulsive interaction between the baryons near the onset. 
We see that this coupling is proportional to the 't Hooft coupling $\lambda$. Interestingly, it gets strong for small asymptotic 
separations $L$ of the flavor branes, measured relative to the maximal separation $\pi/M_{\rm KK}$.

\begin{table}[t]
\centering
\begin{tabular}{c|c|}
\cline{2-2}
&  \multicolumn{1}{|c|}{\multirow{2}{*}{quark density $N_cN_4/V$}} \\ 
& \multicolumn{1}{|c|}{}   \\ \cline{1-2}
\multicolumn{1}{|c|}{\multirow{3}{*}{$B=0$}} &
   \multicolumn{1}{|c|}{\multirow{3}{*}{$\displaystyle{\;\;N_c\,\frac{2(M_q^0)^2}{\frac{3\xi}{2}\lambda\frac{\pi/M_{\rm KK}}{L}}\, (\mu_q - M_q^0)\;\;}$}}      \\ 
\multicolumn{1}{|c|}{}                        &
    \\
\multicolumn{1}{|c|}{}                        &
   \\ \cline{1-2}
\multicolumn{1}{|c|}{\multirow{3}{*}{$\;\;B\rightarrow\infty\;\;$}} &  \multicolumn{1}{|c|}{\multirow{3}{*}{$\displaystyle{N_c\, 
\frac{B}{2\pi^2\tilde{\xi}}\, \left(\mu_q-2M_q^\infty\right)}$}}
    \\
\multicolumn{1}{|c|}{}                        &
    \\ 
\multicolumn{1}{|c|}{}                        &
   \\ \cline{1-2}
\end{tabular}
\caption{Quark density $N_cN_4/V$ induced by wrapped D4-branes close to the baryon onset at vanishing and at asymptotically large magnetic 
field $B$. 
Both expressions depend linearly on the difference between the quark chemical potential and the (effective) constituent quark mass, 
indicating a bosonic nature of the holographic large-$N_c$ baryons discussed here. Since, at nonzero magnetic field, the effective constituent 
quark mass in a baryon is modified by the meson supercurrent, the baryon onset for $B\to\infty$ is at $2M_q^\infty$, not at $M_q^\infty$. 
 For the values of the numerical constants $\xi$, $\tilde{\xi}$ see eqs.\ \eqref{xi} and
\eqref{xitilde}. }
 \label{densitytable}
\end{table}

\subsection{Nonzero magnetic field}
\label{sec:nonzero}

After having checked numerically that baryonic matter, where it exists, has always lower free energy than the mesonic phase, we know that 
the baryon onset
line in the $b$-$\mu$ diagram is defined by $n_4=0$. With the help of eq.\ (\ref{n4}) this gives a condition for $\cosh y_\infty$, into which 
we insert $y_\infty$ from eq.\ (\ref{yinfell}). Together with the condition for the asymptotic separation from eq.\ (\ref{yinfell})
this yields two coupled equations for $\mu$ and $u_c$, 
\begin{subequations}
\bea
\frac{3\mu}{2u_c-3\mu} &=& 
\cosh\,\int_1^\infty \frac{6bu_c u^{3/2}\;du}{\sqrt{4u_c^2[(u^8-1)u_c^3+(u^5-1)b^2]+(u^3-1)b^2\left(2u_c-3\mu\right)^2}}\, ,
\label{onseteq1}\\[2ex]
\frac{u_c^{1/2}\ell}{2} &=& \int_1^\infty
\frac{\sqrt{4u_c^2(u_c^3+b^2)-b^2(2u_c-3\mu)^2}\;du}{u^{3/2}\sqrt{4u_c^2[(u^8-1)u_c^3+(u^5-1)b^2]+(u^3-1)b^2\left(2u_c-3\mu\right)^2}} \, ,
\label{onseteq2} 
\eea
\end{subequations}
where we have changed the integration variable $u\to \frac{u}{u_c}$.
Since the right-hand side of eq.\ (\ref{onseteq1}) is $\ge 1$ we conclude
\be
\frac{u_c}{3}\le \mu_{\rm onset} \le \frac{2u_c}{3} \, .
\ee
(Remember that $u_c$ itself depends on $b$ and $\mu$.) We have seen above that the lower boundary is saturated for $b=0$. 
The upper boundary is saturated for $b\to\infty$, as can be seen from setting $n_4=0$ in eq.\ (\ref{n4}) and using $y_\infty\to\infty$ at 
asymptotically large $b$. Hence, by inserting $\mu=\frac{2u_c}{3}$ into eq.\ (\ref{onseteq2})
we find the asymptotic $u_c$, which in turn gives the asymptotic onset
\be \label{asymp}
\mu_{\rm onset}^\infty = \frac{32\pi}{3\ell^2}\left[\frac{\Gamma\left(\frac{3}{5}\right)}{\Gamma\left(\frac{1}{10}\right)}\right]^2 \simeq 
\frac{0.82}{\ell^2} \, . 
\ee
Closely above that value the behavior of the asymptotic density of ``normal'' baryons $n_4$ is given by (for a derivation see appendix \ref{app2b})
\be\label{n4onsetbinfty}
n_4(b=\infty) = \frac{3b}{2}\frac{\mu-2m_q^\infty}{\tilde{\xi}} + {\cal O}[(\mu-2m_q^\infty)^2]\, ,
\ee
with 
\be \label{xitilde}
\tilde{\xi}\equiv 1-\frac{2}{15\sqrt{\pi}}\frac{\Gamma\left(\frac{1}{10}\right)}{\Gamma\left(\frac{3}{5}\right)} \simeq 0.5194 \, .
\ee
The corresponding dimensionful quark density is shown in the second row of table~\ref{densitytable}. (Remember that the total baryon density $n$
also receives a contribution from the supercurrent, $n=\frac{3}{2}b\jmath + n_4$ with $\jmath(b=\infty) = u_c/3$.) 
Interestingly, at asymptotically large magnetic fields the explicit dependence on the model specific constants $L$, $\lambda$ and $M_{\rm KK}$ 
drops out. This kind of ``universality'' has already been observed in the phases without baryons, see for instance eq.~(4.9) in \cite{Preis:2010cq}.

\begin{figure} [t]
\begin{center}
\includegraphics[width=0.425\textwidth]{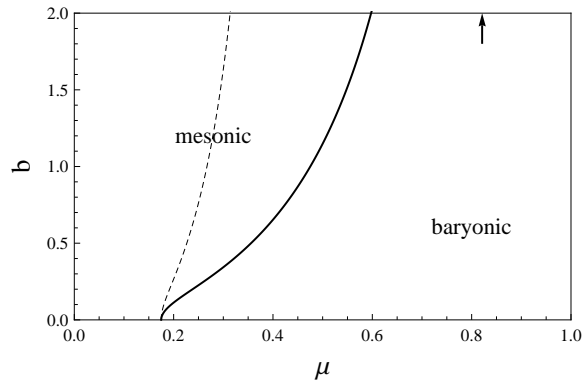}
\caption{Onset of baryonic matter (solid line) in the plane of quark chemical potential $\mu$ and magnetic field $b$.  For comparison, the dashed 
line indicates the baryon onset in the unphysical case of a 
vanishing 
meson supercurrent. In this case, the onset occurs at $\frac{u_c}{3}$ for all $b$. Due to the meson supercurrent, the effective baryon mass is more 
complicated and the 
onset is somewhere between $\frac{u_c}{3}$ ($b=0$) and $\frac{2u_c}{3}$ ($b=\infty$). The arrow indicates the asymptotic
value of the onset line given in eq.\ (\ref{asymp}). In the present model, the baryon onset is always a second-order
phase transition, in contrast to real nuclear matter. For now, we have ignored the chirally restored 
phase. We shall see in sec.\ \ref{sec:phasediagram} that for sufficiently large magnetic field, $b\gtrsim 0.25$,  the transition to baryonic 
matter is replaced by a transition to chirally symmetric quark matter.}
\label{figonset}
\end{center}
\end{figure}

One might have thought that the baryon onset always happens at $\mu = \frac{u_c}{3}$ since $N_c \frac{u_c}{3}$ is the baryon 
mass given by the action of the D4-branes. To understand why the result deviates from this expectation, it is instructive to consider the 
(unphysical) case of an isotropic meson condensate where the meson supercurrent vanishes, 
$\jmath=0$. In this ``cleaner'' case,  the true vacuum with zero pressure $P=0$ and zero baryon density $n=0$ is the ground 
state below the baryon onset. 
Now, eq.\ (\ref{n41}) shows that the onset indeed occurs at $\mu=\frac{u_c}{3}$ for arbitrary values of the magnetic field. 
As a consequence, the chemical potential equals the energy per baryon $\epsilon/n$ at the onset 
(this follows from the thermodynamic relation $P=-\epsilon + \mu n $ and $P=0$ at the onset). The situation with a nonzero supercurrent 
$\jmath$ is different. Here the pressure and the baryon density become nonzero as soon as we switch on $\mu$, and $\mu$ is always larger 
than $\epsilon/n$. Then, at some point, $\mu$ is large enough to support the baryon number induced by $\jmath$ {\it and} by ``normal'' baryons.
This costs more energy than having only ``normal'' baryons and thus the onset happens later than for $\jmath=0$. 
We compare both onset lines, the unphysical one from $\jmath=0$ (dashed) and the physical one with $\jmath\neq 0$ minimizing the free energy 
(solid), in fig.\ \ref{figonset}.

The difference between the two lines can also be thought of as follows. Between the dashed and solid 
lines the system might ``think about'' adding baryons because there is enough energy available from the chemical potential. If the system decided 
to do so, it would 
at the same time have to decrease the supercurrent
discontinuously to have enough energy available for the baryons. (The dashed line is the extreme case where it would have to force the supercurrent 
to zero.) This, however, would lead to a decrease in the total baryon density upon increasing the chemical potential, which is a 
thermodynamically unstable situation. (At the dashed line, the baryon density would jump to zero because the energy is just enough to start adding 
baryons\footnote{\label{burikham}
This unphysical onset is considered in ref.\ \cite{Burikham:2011rg}: the $T=0$ onset in fig.\ 9 of this reference corresponds to 
the dashed line in our fig.\ \ref{figonset}. It seems that this discrepancy originates from incorrectly scaling $S_{\rm D4}$ with the total density
$n$, not $n_4$, see eq.\ (2.4) in ref.\ \cite{Burikham:2011ng} which apparently is used in ref.\ \cite{Burikham:2011rg} 
(and in \cite{Burikham:2009yi}). This difference is very important for the topology of the resulting phase diagram: as we shall see in 
sec.\ \ref{sec:results}, the physical onset line intersects the chiral phase transition line, 
see fig.\ \ref{figPD}; this is not the case for the dashed line.}.)
For any given $b$ the 
solid line marks the ``earliest'' possible point where baryons can be put into the 
system without such a thermodynamic instability.

What can we learn about real-world baryonic matter from these results? We already know that the nature of the onset is different due
to the lack of binding energy. But why do our holographic baryons effectively become heavier in a magnetic field? Can we draw any conclusions from this
observation? Within our model, the increasing critical chemical potential has two reasons. First, as just discussed, it is the meson gradient which is responsible 
for increasing the baryon onset from $\frac{u_c}{3}$ to $\frac{2u_c}{3}$. Second, $u_c$ itself increases with the magnetic field. In the mesonic 
phase $u_c$ can be interpreted as a constituent quark mass \cite{Aharony:2006da,Johnson:2008vna}. In the chiral limit of vanishing bare quark 
masses, it is the chiral condensate which induces such a mass. Therefore, the increase of $u_c$ in a magnetic field suggests an increase
of the chiral condensate and thus is a manifestation of magnetic catalysis (MC).
The constituent quark mass in a baryon is different from the constituent mass in the mesonic phase, $\frac{u_c}{3}$ vs.\ $u_c$, but both
are proportional to $u_c$. Consequently, MC seems to be responsible for the heaviness of magnetized baryons.

\section{Baryon onset in a relativistic mean-field model}
\label{sec:walecka}

In this section we employ the Walecka model \cite{Walecka:1974qa} at zero temperature in a background magnetic field. The Walecka model is a 
relativistic model for dense nuclear matter, where nucleons (or, in extensions of the model, hyperons) interact via Yukawa exchange of mesons. 
In the simplest, isospin-symmetric, version considered here, nucleons interact through the scalar sigma meson and the vector omega meson. This is 
sufficient to model the realistic nucleon-nucleon interaction which is known to have a repulsive short-range (omega) and an attractive intermediate and long-range (sigma) part. Nuclear 
matter being stable at zero pressure, 
having a finite binding energy $E_0\simeq -16.3\,{\rm MeV}$, a saturation density $n_0\simeq 0.153\,{\rm fm}^{-3}$, and showing a first-order 
liquid-gas phase 
transition are all manifestations of this simple but crucial property of the interaction. Beyond the saturation density, the properties of 
nuclear matter are poorly known, and the Walecka model is only one of many competing models. However, for our comparison to holographic baryonic 
matter, we are primarily interested in the baryon onset, and we know that for this purpose -- 
at least for vanishing magnetic field --
the Walecka model describes, by construction, real-world nuclear matter. 

\subsection{Lagrangian}
\label{sec:lagrangian}

The following setup is essentially taken from works where dense nuclear matter
in a magnetic field has been considered in the astrophysical context, see refs.\ \cite{Broderick:2000pe,Broderick:2001qw,Rabhi:2011ej} and
references therein. In neutron stars, the simplest version of baryonic matter consists of 
neutrons, protons, and electrons. Their various chemical potentials are related through the conditions of beta equilibrium and (global) 
electric charge neutrality. For our purpose, these complications are irrelevant since for a comparison to our holographic results 
we are interested in the behavior of a single 
baryon species with a given electric charge and a single baryon chemical potential in a background magnetic field. Therefore, our results will 
be simpler and more transparent, albeit less realistic, compared to the results for astrophysical nuclear matter. 

We start from the Lagrangian 
\be
{\cal L} = {\cal L}_B + {\cal L}_I + {\cal L}_M  \, , 
\ee
containing a baryonic part ${\cal L}_B$, an interaction part ${\cal L}_I$, and a mesonic part ${\cal L}_M$. 
The baryonic part is
\be
{\cal L}_B = \bar{\psi}\left(i\gamma^\mu D_\mu-m_B+\mu_B\gamma^0-\frac{1}{2}\kappa\sigma_{\mu\nu}F^{\mu\nu}\right)\psi \, , 
\ee
with the baryon mass $m_B$ and the baryon chemical potential $\mu_B$. 
The baryons feel the magnetic field through the covariant derivative  
$D_\mu = \partial_\mu+iqA_\mu$ with the baryon electric charge $q$ and the electromagnetic gauge field $A_\mu$,
which encodes the background magnetic field in the $x_3$-direction, $A_\mu = (0,x_2B,0,0)$. In addition, baryons have an anomalous magnetic moment 
$\kappa$ whose effect is included by a magnetic dipole term with $\sigma_{\mu\nu}=\frac{i}{2}[\gamma_\mu,\gamma_\nu]$ and the 
field strength tensor $F^{\mu\nu}=\partial^\mu A^\nu - \partial^\nu A^\mu$. It is important to keep in mind that this term is obviously an effective,
not a fundamental, way to take into account the anomalous magnetic moment. In particular, the present approach can not be trusted for
arbitrarily large magnetic fields, as we shall see more explicitly below.

The interaction term consists of two Yukawa contributions for the $\sigma$ and the $\omega$,
\be
{\cal L}_I = g_\sigma \bar{\psi}\sigma\psi - g_\omega\bar{\psi}\gamma^\mu\omega_\mu\psi \, , 
\ee
with coupling constants $g_\sigma, g_\omega>0$.  
The mesonic part includes cubic and quartic scalar self-interactions,
\be \label{LM}
{\cal L}_M = \frac{1}{2}(\partial_\mu\sigma\partial^\mu\sigma-m_\sigma^2\sigma^2) -\frac{1}{4}\Omega^{\mu\nu}\Omega_{\mu\nu}+\frac{1}{2}m_\omega^2
\omega^\mu\omega_\mu - \frac{b}{3}m_B(g_\sigma\sigma)^3-\frac{c}{4}(g_\sigma\sigma)^4 \, , 
\ee
with $\Omega^{\mu\nu} = \partial^\mu \omega^\nu - \partial^\nu \omega^\mu$ and the sigma and omega masses $m_\sigma$ and $m_\omega$.
With the given values $m_B=939\,{\rm MeV}$, $m_\omega = 783\,{\rm MeV}$, $m_\sigma\sim 550\,{\rm MeV}$, the model has four free parameters which 
are fitted to reproduce the saturation density, the binding energy, the compressibility, and the Landau mass at saturation (all for $B=0$), 
which gives $\frac{g_\sigma^2}{4\pi}=6.003$, $\frac{g_\omega^2}{4\pi}=5.948$, $b=7.950\times 10^{-3}$, and $c=6.952\times 10^{-4}$. 
We shall employ the mean-field approximation where the meson fluctuations are neglected, and the meson condensates $\bar{\sigma}$ and 
$\bar{\omega}_0$ have to be determined from minimization of the free energy. The basic equations of the model in this approximation 
are as follows (see for instance ref.\ \cite{Schmitt:2010pn} for a pedagogical derivation). The pressure is
\be \label{P}
P = -\frac{1}{2}m_\sigma^2\bar{\sigma}^2 -\frac{b}{3}m_B(g_\sigma \bar{\sigma})^3-\frac{c}{4}(g_\sigma\bar{\sigma})^4+
\frac{1}{2}m_\omega^2\bar{\omega}_0^2+P_B \, ,
\ee
where, at zero magnetic field, the renormalized baryonic pressure is given by 
\be \label{PB}
P_B = 2T\int\frac{d^3{\bf k}}{(2\pi)^3}\,\ln\left[1+e^{-(\epsilon_k-\mu_*)/T}\right] \, . 
\ee
Here, $\mu_* \equiv \mu_B-g_\omega\bar{\omega}_0$ plays the role of the Fermi energy at zero temperature (the chemical potential in the 
thermodynamic sense it still 
$\mu_B$, not $\mu_*$). The baryon dispersion is $\epsilon_k = \sqrt{k^2+m_*^2}$ with an effective baryon mass $m_*\equiv m_B-g_\sigma\bar{\sigma}$.
The stationarity equations for the meson condensates are
\begin{subequations} \label{mstarom}
\bea
m_* & = & m_B-\frac{g_\sigma^2}{m_\sigma^2}n_s+\frac{g_\sigma^2}{m_\sigma^2}\left[bm_B(m_B-m_*)^2
+c(m_B-m_*)^3\right] \, , \label{mstar}\\
\bar{\omega}_0 &=& \frac{g_\omega}{m_\omega^2}n_B \, , 
\eea 
\end{subequations}
where $n_B=\frac{\partial P_B}{\partial \mu_B}$ and $n_s=-\frac{\partial P_B}{\partial m_B}$ are the baryon and scalar densities, respectively.

Next we include the magnetic field, wherefore we need to distinguish between charged and neutral baryons. 
For our purpose the only relevant effect of the magnetic field concerns the single-baryon dispersion relations, which can be found in  
ref.\ \cite{Rabhi:2011ej}.

\subsection{Charged baryons}
\label{sec:charged}

Our reference example of charged baryonic matter is pure proton matter with charge $q=+e$ and an anomalous magnetic moment $\kappa = 1.79\,\mu_N$ 
where $\mu_N = \frac{e}{2m_p} \simeq 3.15\times 10^{-18}\,{\rm MeV}\,{\rm G}^{-1}$ is the nuclear magneton. We shall, however, vary
the anomalous magnetic moment below to study its effect on the baryon onset. Remember that we ignore any neutrality constraint 
and Coulomb effects. For charged baryons the dispersion $\epsilon_k$ has to be replaced by
\be \label{epsiloncharged}
\epsilon_{k_\parallel,\nu,s} = \sqrt{k_\parallel^2+M_{\nu,s}^2} \, , \qquad M_{\nu,s} \equiv \sqrt{m_*^2+2\nu|q|B}-s\kappa B \, ,
\ee
with $k_\parallel$ being the momentum in the direction of the magnetic field, $s=\pm 1$, and $\nu=n+(1-s\,{\rm sgn}\,q)/2$ with $n=0,1,2,\ldots$, 
such that for positive charge we have $\nu = 0,1,2,\ldots$ for $s=+1$ and $\nu = 1,2,3,\ldots$ for $s=-1$, and vice versa for negative charge. 
This means that the lowest Landau level (LLL)
$\nu=0$ is only populated by $s=+1$ fermions for $q>0$ and $s=-1$ fermions for $q<0$, while both $s=+1$ and $s=-1$ 
contribute to all higher Landau levels. The three-dimensional 
momentum integral has to be replaced by a one-dimensional integral over $k_\parallel$ and a sum over Landau levels and spin degrees of freedom,
\be
2\int\frac{d^3{\bf k}}{(2\pi)^3} \to \frac{|q|B}{2\pi^2}\sum_{s=\pm}\sum_\nu\int_0^\infty dk_\parallel \, . 
\ee
With these replacements in the baryonic pressure (\ref{PB}) we obtain at zero temperature
\be \label{Pcharged}
P_B= \frac{|q|B}{4\pi^2}\sum_{s=\pm}\sum_\nu^{\nu_{\rm max}}\left(\mu_*k_{F,\nu,s}-M_{\nu,s}^2\ln\frac{k_{F,\nu,s}+\mu_*}{M_{\nu,s}}\right) \, ,
\ee
and 
\begin{subequations} \label{nBnscharged}
\bea
n_B&=&\frac{|q|B}{2\pi^2}\sum_{s=\pm}\sum_\nu^{\nu_{\rm max}} k_{F,\nu,s} \, , \\
n_s&=& \frac{|q|Bm_*}{2\pi^2}\sum_{s=\pm}\sum_\nu^{\nu_{\rm max}}\frac{M_{\nu,s}}{\sqrt{m_*^2+2\nu|q|B}}\,\ln\frac{k_{F,\nu,s}+\mu_*}{M_{\nu,s}} 
\, ,
\eea
\end{subequations}
where $k_{F,\nu,s}=\sqrt{\mu_*^2-M_{\nu,s}^2}$ is the longitudinal Fermi momentum, and where the Landau levels are occupied
up to $\nu_{\rm max} = \left\lfloor\frac{(\mu_*+s\kappa B)^2-m_*^2}{2|q|B}\right\rfloor$. By inserting eqs.\ (\ref{nBnscharged}) into 
eqs.\ (\ref{mstarom}) and solving the resulting equations for $m_*$, $\bar{\omega}_0$, one can 
compute the thermodynamic properties for arbitrary $\mu_B$ and $B$. Since we are only interested in the baryon onset, we have the
additional condition $P=0$, where $P$ is obtained by inserting eq.\ (\ref{Pcharged}) into eq.\ (\ref{P}). Hence we have a system of three 
equations to be solved for $m_*$, $\bar{\omega}_0$, and $\mu_B$ at any given $B$.

\begin{figure} [t]
\begin{center}
\hbox{\includegraphics[width=0.45\textwidth]{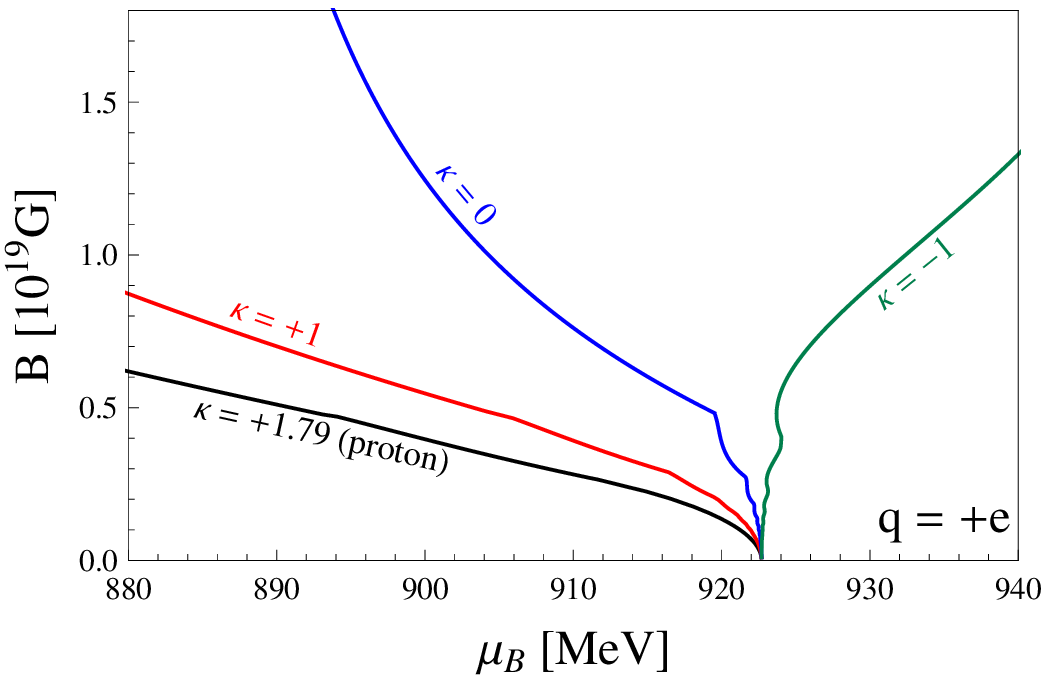}\hspace{0.3cm}
\includegraphics[width=0.45\textwidth]{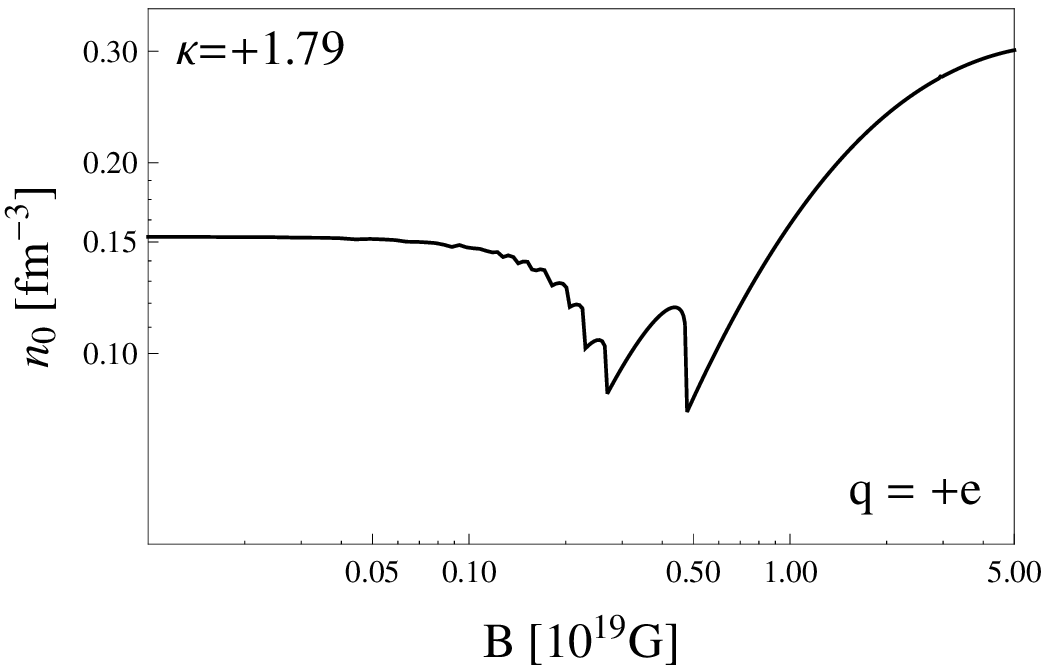}}
\caption{Left panel: zero-temperature transition in the plane of baryon chemical potential $\mu_B$ and magnetic field $B$ 
from the vacuum (to the left of each line) to baryonic matter (to the right of each line) for charged baryons with $q=+e$ and various values of the 
anomalous magnetic moment $\kappa$ (in units of the nuclear magneton $\mu_N$). Each line represents a first-order phase transition since the 
baryon density jumps from zero to a finite value $n_0$.  Right panel: corresponding baryon density $n_0$ along the onset line 
on a doubly logarithmic plot for the case 
of pure proton matter, $\kappa=+1.79$ (the other three cases from the left plot would lead to similar curves). 
The oscillations are due to successive occupation of 
Landau levels. For $B\gtrsim 0.5\times 10^{19}\, {\rm G}$
only the LLL is occupied.}
\label{figcharged}
\end{center}
\end{figure}

The result for pure proton matter and three more (unphysical) types of charged baryonic matter with $q=+e$, distinguished by different values 
for $\kappa$, is shown in fig.\ \ref{figcharged} where we plot the onset lines and, for proton matter, the baryon density along the onset. 
(In order to reproduce
the nuclear ground state density $n_0$ at $B=0$ we have multiplied the pressure by a factor of 2. In other words, we have started from the 
isospin-symmetric, $B=0$ Walecka model for protons and neutrons and have added the effect of the magnetic field as if both nuclear  
species had the same charge and anomalous magnetic moment.)  

The $B=0$ onset occurs at $\mu_B = m_B+E_0 \simeq 922.7 \, {\rm MeV}$ where $E_0\simeq -16.3\,{\rm MeV}$ is the nuclear binding energy. 
Magnetic fields of the order of $10^{18}\, {\rm G}$ and larger change this significantly, in accordance with the simple estimate in 
sec.\ \ref{sec:context}. 
We observe oscillations due to the Landau level structure in the onset line as well as in the 
onset density. At sufficiently large magnetic fields only the LLL is occupied. (If there was no binding energy, the density at the onset
would be infinitesimally small, and all along the onset line the LLL would be the only relevant state. In other words, only due to the presence of 
a finite binding energy the onset line shows a behavior reminiscent of de Haas-van Alphen oscillations at small magnetic 
fields.\footnote{The higher Landau levels in fact induce cusps in some of the onset lines, see for instance the case 
$\kappa=0$ where $\mu_B^{\rm cusp} \simeq 920\,{\rm MeV}$. In a small vicinity 
around this cusp there are, for a given $B$, two solutions for the onset: one where only the LLL becomes occupied, and one where the lowest and 
the first Landau level become occupied simultaneously. The resulting two onset lines intersect, and for any given 
$B$ the line where the onset occurs at a lower 
$\mu_B$ is the physical one. The reason is that the ``later'' onset would keep the system in the vacuum state $P=0$, in a region where 
the ``earlier'' onset already leads to $P>0$. (For $\mu_B<\mu_B^{\rm cusp}$, the LLL onset is ``earlier'', for $\mu_B>\mu_B^{\rm cusp}$
it is ``later'', hence the cusp.)})
Since we have, without loss of generality, fixed the 
electric charge to be positive, the LLL is occupied by $s=+1$ baryons in all four cases considered here. Note that in one 
of the four cases shown in the figure, the anomalous magnetic moment $\kappa$ is negative. Since $\kappa<0$ favors $s=-1$ baryons, it can in 
principle be less costly to put baryons in the $\nu=1$, $s=-1$ state than in the $\nu=0$, $s=+1$ state such that for sufficiently 
large magnetic fields all baryons sit in the first, not in the lowest, Landau level. This does not occur for the cases shown here, but a 
simple estimate shows that it occurs if the anomalous moment and the charge have opposite sign and $|\kappa|$ becomes of the order of or larger 
than the modulus of the normal magnetic moment $\frac{|q|}{2m_B}$. 

One can check that for large magnetic fields the onset lines are well approximated by the curves $\mu_*=m_*-\kappa B$, which turn out to become 
straight lines for large $B$. However, we need to remember that we have used an effective approach for the anomalous magnetic moment. As a 
consequence, we should 
not trust our results for $\kappa B$ becoming of the order of or larger than $m_*$. In that case, as we see for 
instance from eq.\ (\ref{epsiloncharged}), $M_{\nu=0,s=+1}$  would become negative (for $\kappa>0$). In the case of proton matter we find 
$\kappa B \simeq 0.4\,m_*$ at $B=5\times 10^{19}{\rm G}$, which suggests that our approach is still reliable in the plotted range.
We expect that in a full treatment the onset lines saturate 
for asymptotic values of $B$. This is the case for $\kappa=0$, where our approach can be used for arbitrarily large $B$. 

\subsection{Neutral baryons}
\label{sec:neutral}

\begin{figure} [t]
\begin{center}
\hbox{\includegraphics[width=0.45\textwidth]{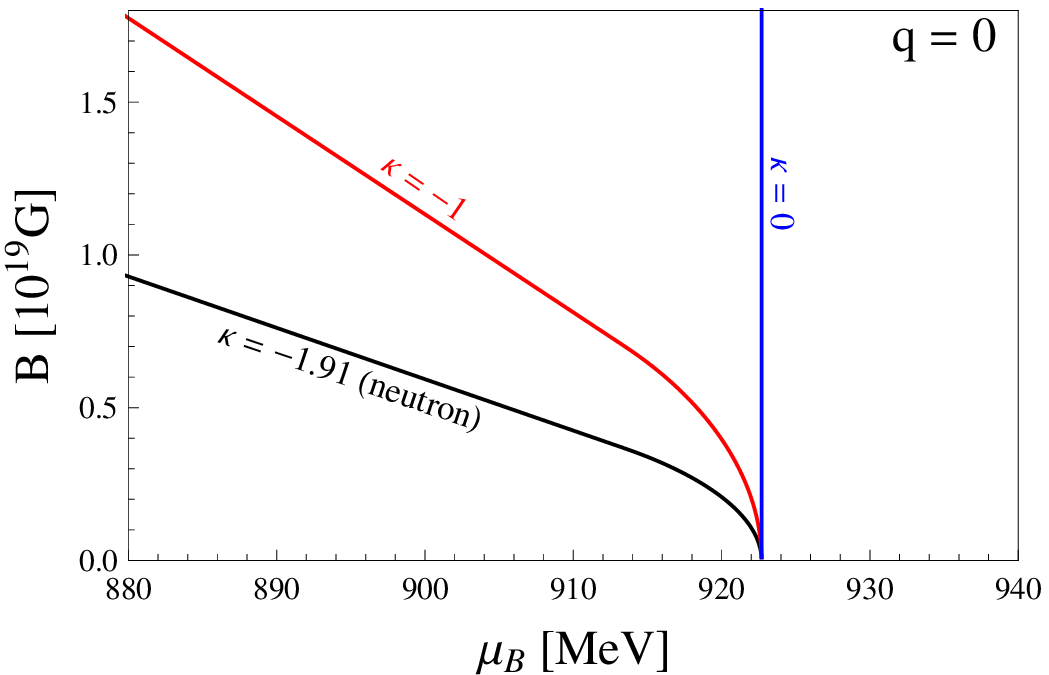}\hspace{0.3cm}
\includegraphics[width=0.45\textwidth]{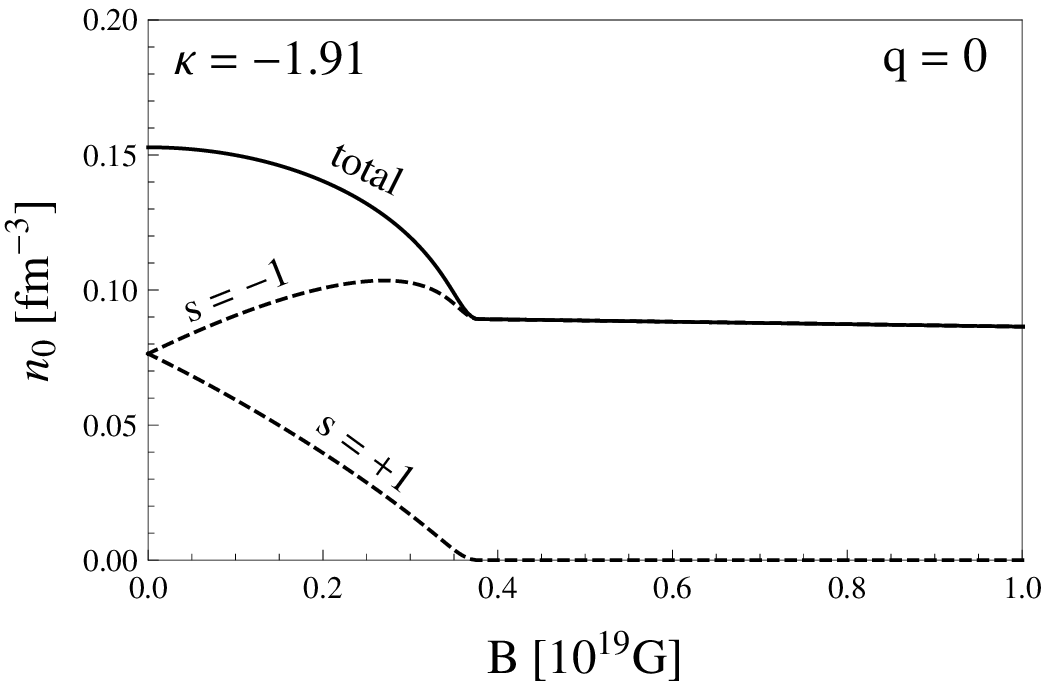}}
\caption{Same as fig.\ \ref{figcharged}, but for neutral baryons, $q=0$. Due to the 
symmetry under $\kappa \to -\kappa$, we can restrict ourselves to negative $\kappa$. The density along the onset in the right panel
is shown for pure neutron matter. We have plotted the total baryon density (solid) and the contributions from the $s=+1$ and $s=-1$ states (dashed).
We see that at $B\gtrsim 0.4 \times 10^{19}\,{\rm G}$ the system is fully polarized in the $s=-1$ state.}
\label{figneutral}
\end{center}
\end{figure}
For the case of neutral baryons we may think of pure neutron matter with $q=0$ and $\kappa = -1.91\,\mu_N$. In this case there is no Landau 
quantization and the single-baryon excitations become
\be \label{epsilonneutral}
\epsilon_{{\bf k},s} = \sqrt{k_\parallel^2+\left[\sqrt{m_*^2+k_\perp^2}-s\kappa B\right]^2} \, ,
\ee
where ${\bf k}_\perp$ is the momentum perpendicular to the magnetic field. Replacing the momentum integral in the nucleonic pressure (\ref{PB}),
\be
2\int\frac{d^3{\bf k}}{(2\pi)^3} \to \frac{1}{2\pi^2}\sum_{s=\pm}\int_0^\infty dk_\perp k_\perp\int_0^\infty dk_\parallel  \, , 
\ee
and performing the integrals at zero temperature yields 
\bea
P_B&=& \sum_{s=\pm}\frac{\Theta(\mu_*+s\kappa B-m_*)}{2\pi^2} 
\left[\mu_* k_{F,s}\left(\frac{k_{F,s}^2}{12}-\frac{M_s^2}{8}-\frac{s\kappa BM_s}{3}\right)\right. \non[2ex]
&& \left. \hspace{1cm} +\, \frac{s\kappa B \mu_*^3}{6} \arccos\frac{M_s}{\mu_*} 
+ M_s^3\left(\frac{M_s}{8}+\frac{s\kappa B}{6}\right)\ln\frac{\mu_*+k_{F,s}}{M_s}\right] \, ,
\eea
with $M_s \equiv  m_*-s\kappa B$ and the longitudinal Fermi momentum $k_{F,s} = \sqrt{\mu_*^2-M_s^2}$. 
The baryon and scalar densities become 
\begin{subequations}
\bea
n_B&=&\sum_{s=\pm}\frac{\Theta(\mu_*+s\kappa B-m_*)}{2\pi^2} \left[\frac{k_{F,s}^3}{3}+\frac{s\kappa B}{2}
\left(\mu_*^2\arccos\frac{M_s}{\mu_*}-k_{F,s} M_s \right)\right] \, , \\[2ex]
n_s&=&m_*\sum_{s=\pm}\frac{\Theta(\mu_*+s\kappa B-m_*)}{4\pi^2}\left(\mu_*k_{F,s}-M_s^2\ln\frac{\mu_*+k_{F,s}}{M_s}\right) \, .
\eea
\end{subequations}
We observe that the system is symmetric under $\kappa\to -\kappa$ because after a sign change of $\kappa$ the contributions from $s=+1$ and $s=-1$
have simply exchanged their roles. Analogously to the charged case we can now compute the baryon onset. 
The results are shown in fig.\ \ref{figneutral}. For large magnetic fields, the onset curves can be approximated by the simple straight lines 
$\mu_B=m_B-|\kappa|B$. 
In this regime the system is fully polarized, i.e., the baryons are all in the $s=-1$ ($s=+1$) state if $\kappa <0$ ($\kappa>0$), as can be seen 
in the right panel of the figure. Having in mind the symmetry under $\kappa\to -\kappa$ it is instructive to go back to the 
results for charged baryons. In the left panel
of fig.\ \ref{figcharged}, the onset lines for $\kappa=+\mu_N$ and $\kappa = -\mu_N$ are obviously far from identical. This 
is a LLL effect: in the LLL the baryons with $q>0$ become heavier for $\kappa <0$ and lighter for $\kappa>0$. If we let $q\to 0$ the higher Landau levels
become more and more important, and the two curves indeed approach each other and become identical for $q=0$, see onset line for $\kappa=-\mu_N$
in fig.\ \ref{figneutral}. Now the baryons in the lowest energy state become lighter for either sign of $\kappa$.

\subsection{Binding energy}

\begin{figure} [t]
\begin{center}
\includegraphics[width=0.45\textwidth]{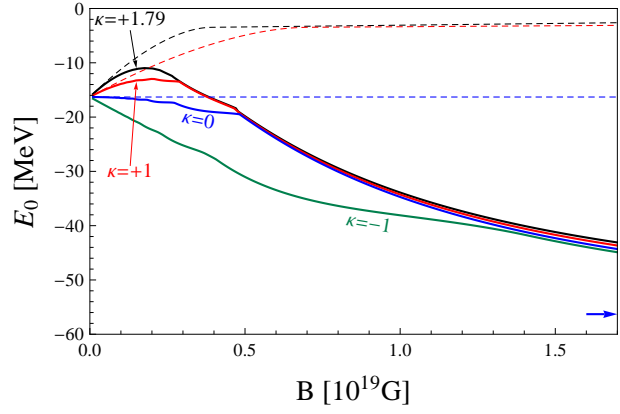}
\caption{Solid lines: binding energy $E_0$ along the baryon onset for charged baryons with charge $q=+e$ and various 
anomalous magnetic moments $\kappa$, corresponding to the four cases shown in fig.\ \ref{figcharged}. Dashed lines: binding energies for 
baryons with the same values for $\kappa$, but $q=0$. They approximate the curves for the charged baryons at small $B$
where many Landau levels are occupied. The arrow at $E_0=-56.3\,{\rm MeV}$ indicates the asymptotic value of the binding energy for 
$\kappa=0$, see eq.~\eqref{bindlargeB} (for $\kappa\neq 0$ our approach does not allow for arbitrarily large magnetic fields).}
\label{figbinding}
\end{center}
\end{figure}
Let us now compare the baryon onset in the holographic model, fig.\ \ref{figonset}, with the ones in the Walecka model for charged and neutral 
baryons, figs.\ \ref{figcharged} and \ref{figneutral}. The first simple observation is that in the holographic case the larger the magnetic field, 
the larger the energy needed to create baryons. This seems to be in contrast to the results from the previous two subsections which suggest that 
a magnetic field tends to make nuclear matter energetically less costly. In order to compare the holographic with the Walecka result
in a sensible way, we need to isolate the effect of the binding energy $E_0$. 

In the absence of a magnetic field the binding energy is given by $E_0 = \frac{E}{A} -m_B$, i.e., by the energy per baryon 
$\frac{E}{A}=\frac{\epsilon}{n_B}$ relative to its mass $m_B$. To be precise, by mass we mean  
the energy that is needed to put a single, non-interacting baryon into the lowest single-particle state of the system. 
To generalize this concept to finite magnetic fields, let us first consider charged baryons. In this case, the single-particle ground 
state energy without nucleon-nucleon interaction is $m_B-\kappa B\,{\rm sgn}\, q$ according to eq.\ (\ref{epsiloncharged}). Thus we define
\be
E_0 = \frac{\epsilon}{n_B} -(m_B-\kappa B\,{\rm sgn}\, q) \, , 
\ee
where, at the onset, $\frac{\epsilon}{n_B}=\mu_B$. We plot $E_0$ as a function of the magnetic 
field in fig.\ \ref{figbinding}.
The plot shows that in the regime where higher Landau levels are occupied, the binding energy depends strongly on the anomalous magnetic 
moment, while it varies very little with $\kappa$ in the LLL regime. For $\kappa=0$ we can take the limit of asymptotically large magnetic 
fields and find that the onset approaches the line $\mu_*=m_*$, hence with the definitions of $\mu_*$ and $m_*$ below eq.~\eqref{PB} we have
\be\label{bindlargeB}
E_0(\kappa=0,B\to\infty) = g_\omega\bar{\omega}_0-g_\sigma\bar{\sigma} \simeq - 56.3\,{\rm MeV}\, .
\ee
Here, $\bar{\omega}_0$ and $\bar{\sigma}$ are complicated functions of the parameters of the model. Nevertheless, this expression is 
very instructive since it shows the effect of the meson condensates in a very transparent way: $E_0$ is negative because the scalar meson 
condensate, responsible for the attractive interaction, becomes sufficiently large compared to the vector meson condensate, which is responsible 
for the repulsive interaction. There is no such simple expression for $B=0$. Interestingly, 
$E_0(B\to\infty)$ is nonvanishing only in the presence of scalar self-interactions. For asymptotically large $B$, one can show 
analytically that after setting $b=c=0$ in eq.\ (\ref{P}) and (\ref{mstar}) we would obtain $\mu_B=m_B$ and thus $E_0=0$. 

As a result of this discussion we conclude that the  magnetic field has two effects: it changes the mass and the binding energy.
For charged baryons whose charge and anomalous magnetic moment have the same sign,
both effects work in favor of creating baryonic matter, at least for sufficiently large magnetic fields: 
first, they decrease the mass $m_B-  \kappa B\,{\rm sgn}\, q$, 
and second, they lead to a larger binding energy $|E_0|$. At smaller magnetic field and/or different signs of $q$ and $\kappa$, things are more 
complicated and can be read off from figs.\ \ref{figcharged} and \ref{figbinding}. 
For neutral baryons, a magnetic field always decreases the mass, $m_B- |\kappa| B$, and baryonic matter is always favored by a nonzero magnetic field.
If we compare charged with neutral baryons at the same anomalous magnetic moment, for instance the curves for 
$\kappa=+\mu_N$ in figs.\ \ref{figcharged} and \ref{figneutral}, we see that neutral baryonic matter is favored a bit less than charged 
one. The reason is that for neutral baryons $|E_0|$ becomes smaller for large magnetic fields, as fig.\ \ref{figbinding} demonstrates.

\subsection{Large $N_c$}

How does this picture change when we go to the large-$N_c$ limit?  One way to answer this question is to simply rescale the parameters
of the model with the appropriate powers of $N_c$, as suggested from large-$N_c$ arguments \cite{Witten:1979kh,McLerran:2007qj}. This has been done
without magnetic field in ref.\ \cite{Bonanno:2011yr}. While the rescaling of most parameters is unambiguous, one has to make a decision
about the generalization of the scalar meson to large values of $N_c$. In the traditional quark-antiquark picture, its mass would scale like
$N_c^0$. However, we may take the heaviness of the lightest scalar meson in the Sakai-Sugimoto model as a hint that this picture is incorrect,
see also refs.\ \cite{Pelaez:2003dy,Pelaez:2004xp,Parganlija:2010fz}.
The alternative
that we shall consider here is that the lightest scalar meson is a tetraquark state \cite{Jaffe:1976ig}. Let us denote the generalization to 
arbitrary values of $N_c$ of this tetraquark state by $\chi$. 
This state is composed of $N_c-1$ quarks and $N_c-1$ antiquarks such that its mass 
scales like $m_\chi\sim 2(N_c-1)\sim N_c$ \cite{Liu:2007tj}. There are other possibilities
for the nature of the scalar meson which we do not consider here. The results of ref.\ \cite{Bonanno:2011yr}, however, 
suggest that the main conclusions of our following discussion do not depend on the detailed nature of this controversial meson state.
Consequently, we rescale \cite{Bonanno:2011yr,Witten:1979kh,McLerran:2007qj}
\be
m_B,m_\chi,\kappa,q \sim N_c \, , \quad m_\omega, g_\chi \sim N_c^0 \, , \quad g_\omega \sim N_c^{1/2}  \, .
\ee
(The $N_c$-dependence of $\kappa$ has been computed in ref.\ \cite{Hashimoto:2008zw}; note that we also assume the charge $q$ to scale with $N_c$).
The $N_c$-dependence of the self-interactions of $\chi$ is given by $b\sim e^{-N_c}$, $c\sim N_c$; this is suggested by the arguments 
explained in ref.\ \cite{Bonanno:2011yr}. 

Let us first briefly discuss the 
scenario without magnetic field. In this case the stationarity equations of the Walecka model (\ref{mstarom}) imply that the Fermi momentum 
$k_F =\sqrt{\mu_*^2-m_*^2}$ scales like $N_c^0$, although both $\mu_*$ and $m_*$ are proportional to $N_c$. Therefore, in the large-$N_c$ limit,
one can expand the pressure for small $k_F \ll m_*, \mu_*$ (which, as an aside, yields $P\sim N_c$). The resulting equation $P=0$ for the baryon 
onset then becomes very simple and yields, together with the stationarity equations the trivial solution $m_*=m_B$, $\mu_*=\mu_B$ and $E_0=0$, i.e., 
the baryon onset becomes a second-order transition. This result is in fact obtained for all $N_c\ge 4$ \cite{Bonanno:2011yr}.

Now we switch on the magnetic field. Our numerical results show that, not surprisingly, the binding energy remains zero, and as a consequence
the baryon onset curves are simply straight lines in the $B$-$\mu_B$ plane for all $N_c\ge 4$: for charged baryons they are given by 
$B = \frac{m_B-\mu_B}{\kappa}$, 
while for neutral baryons $B = \frac{m_B-\mu_B}{|\kappa|}$. Note the slight subtlety related to the order of limits 
$q\to 0$ and $N_c\to\infty$: consider the onset lines of charged baryons at $N_c=3$ for $\kappa=\pm \mu_N$, see fig.\ \ref{figcharged}. These are two 
very different lines. As discussed above, these lines merge for $q\to 0$. They remain on top of each other if now we let $N_c\to\infty$. 
On the other hand, if we first take the limit $N_c\to \infty$ they remain separated and become linear with opposite slopes. Now letting $q\to 0$ 
does not 
change the result. Hence the symmetry of the onset magnetic field under $\kappa\to -\kappa$ if we take the large-$N_c$ limit of our neutral baryons 
and the {\it anti}-symmetry for the charged baryons.  

With these preparations, what can we learn from our holographic result? We know that there is no binding energy in this case, and  
the onset line simply indicates the $B$-dependent baryon mass (at least without meson supercurrent, otherwise the situation is 
more complicated, as discussed in sec.\ \ref{sec:nonzero}). This mass gets larger with increasing magnetic field. 
In the Walecka model, the only case with increasing mass is the case 
of charged baryons with $q$ and $\kappa$ having opposite sign. Since in the present context a nonzero electric charge is equivalent to 
the existence of Landau levels, it is interesting to ask whether there is any Landau level structure for our holographic baryons. We know that 
in the chirally restored phase of the Sakai-Sugimoto model there is indeed a phase
transition which has similar properties as a transition into the LLL \cite{Preis:2010cq,Lifschytz:2009sz}. In terms of the solution to the 
equations of motion, the ``LLL'' phase corresponds to a trivial solution, $z_\infty = \infty$ in the notation of ref.\ \cite{Preis:2010cq}.
This solution always exists, but in some regions of the parameter space is disfavored compared to a nontrivial solution, interpreted as
a phase of higher Landau levels.  In the baryonic phase, there is also a trivial solution, namely $y_\infty = 0$, see eqs.\ (\ref{yinfell}). 
However, this solution is unphysical because it leads to an infinite baryon density and 
thus infinite free energy. Although we find several nontrivial solutions in certain parameter regions, the solution which is continuously connected 
to the solution at the onset is always the energetically preferred one. Hence there is no phase transition within the baryonic phase.
This is consistent with the apparent bosonic nature of the holographic large-$N_c$ baryons (see sec.\ \ref{sec:zeroB}) because for a Bose condensate at zero temperature we do not expect de Haas-van Alphen oscillations. 

Besides the overall tendency of the $B$-dependent mass, how about its linear behavior seen in the large-$N_c$ limit of the Walecka model? 
The holographic results
show an approximately linear behavior only for intermediate magnetic fields. For large fields, a comparison to the Walecka results would 
only be sensible if the latter included a more elaborate treatment of the anomalous magnetic moment. But also at small magnetic fields our 
holographic onset line differs from the linear behavior, 
in fact we find $\mu_{\rm onset} = m_q^0 + {\rm const} \times b^2 + \ldots$. It would be interesting to compute the onset in other 
field-theoretical approaches. After all, the Walecka 
model does not know about dynamical chiral symmetry breaking and thus cannot be expected to show effects from MC and the meson supercurrent, 
which, as mentioned above, 
seem to be the driving forces for the $B$-dependence of our effective holographic
baryon mass. On the other hand, also our holographic approach should be refined for a more meaningful comparison. In particular, 
a generalization to two flavors is necessary to describe realistic nuclear matter.

\section{Phase diagram with holographic baryonic matter}
\label{sec:phasediagram}

We now come back to the Sakai-Sugimoto model. 
In this section we are mainly interested in the chiral phase transition in the presence of baryonic matter. This requires the inclusion
of chirally symmetric matter into our analysis.  Since
we consider both broken and symmetric phases in the deconfined background, the metric does not change. However, the embedding of the 
flavor branes into the background is different. In the Sakai-Sugimoto model, chirally symmetric matter is described by a configuration where 
the D8- and $\overline{\rm D8}$-branes are disconnected and have a trivial embedding in the background geometry, $x_4'(u)=0$. As in 
the chirally broken phase, we shall work in the approximation $f(u)\simeq 1$. In contrast to the broken phase (see discussion at the beginning of 
sec.\ \ref{sec:eom}) this approximation corresponds unambiguously to the zero-temperature limit. The reason is that, since the flavor branes do 
not connect, the holographic direction on these branes is parameterized by $u\in [u_T,\infty]$. Hence, $f(u)=1-\frac{u_T^3}{u^3}$ can only
be approximated by 1 if $u_T \to 0$, i.e., $T \to 0$ (see sec.~4 of
ref.~\cite{Preis:2010cq} as well as ref.~\cite{Lifschytz:2009sz} for the full temperature-dependent treatment and the limit $T\to0$). 

Our main result of this section will thus be the phase diagram in the $b$-$\mu$ plane at
zero temperature. For our purpose this is not a severe restriction for the following reasons. Judging from our previous results without baryonic matter, the most interesting physics is observed at $T=0$. For instance, IMC is most pronounced at $T=0$. The main effect of nonzero temperature is to disfavor the chirally broken phase, and the non-monotonic behavior of the $T=0$ chiral phase transition line is weakened, but not altered in a nontrivial way. Moreover, since our main results concern matter at large chemical potential, the only application will be in neutron stars where the temperature is negligibly small compared to the chemical potential. 

\subsection{Results}
\label{sec:results}

\begin{figure} [t]
\centerline{ignoring baryons \hspace{6.5cm} including baryons \hspace{1cm}}
\begin{center}
\hbox{\includegraphics[width=0.47\textwidth]{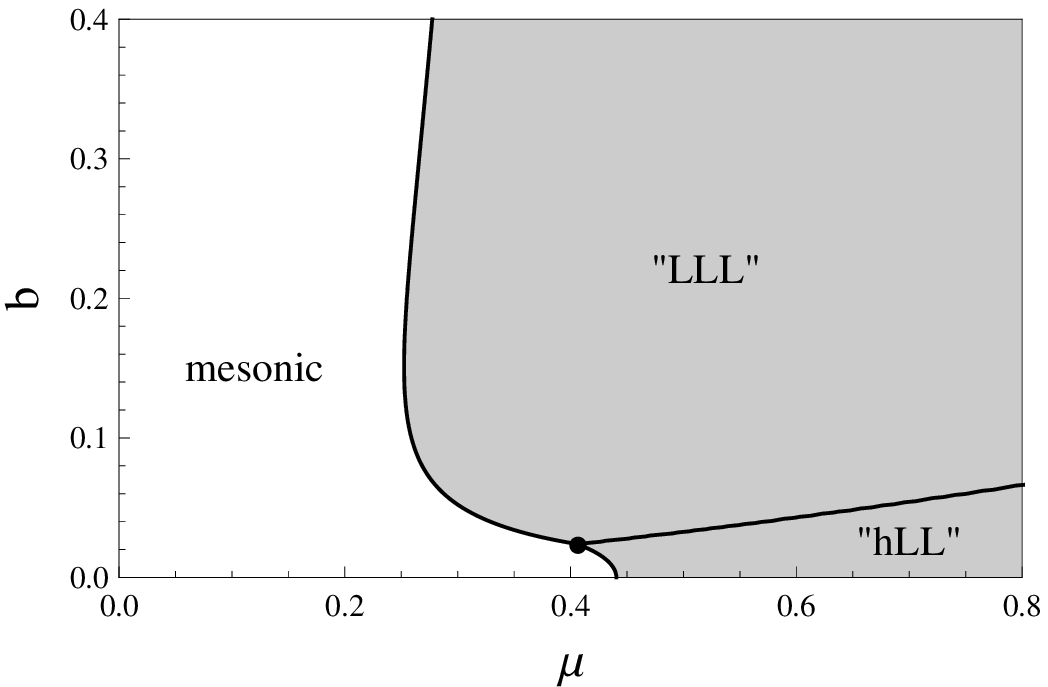}
\includegraphics[width=0.47\textwidth]{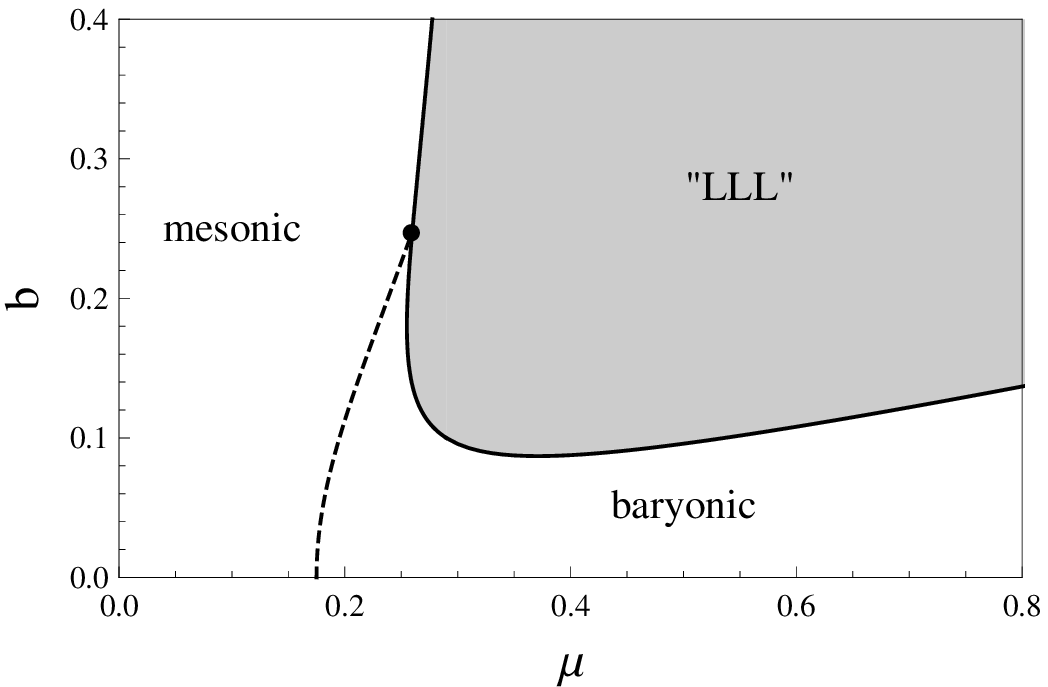}}
\caption{Zero-temperature phase diagram in the plane of magnetic field $b$ and quark chemical potential $\mu$ without (left) and with (right)
baryonic matter in the deconfined geometry of the Sakai-Sugimoto model. In shaded areas the ground state is chirally symmetric, while in unshaded
areas chiral symmetry is spontaneously broken. Solid lines are first-order phase transitions, the dashed line is the baryon onset. The two 
different chirally symmetric phases are reminiscent of phases where only the lowest Landau level (``LLL'') and where higher 
Landau levels (``hLL'') are occupied. 
Recall that $b$ and $\mu$ are dimensionless quantities that involve dimensionful parameters of the model. As a rough estimate,
$b=0.1$ corresponds to $|q|B \sim 5\times 10^{18}\, {\rm G}$. (This estimate is obtained by using the simple fit of ref.\ \cite{Preis:2010cq}
which involves the assumption that the $b=0$ chiral phase transition without baryonic matter at $\mu\approx 0.44$ occurs at a quark chemical potential of 
$400\, {\rm MeV}$.)}
\label{figPD}
\end{center}
\end{figure}

For the phase diagram, we have to compare the free energies of three phases, corresponding to three different embeddings of the flavor branes, 
$\vee$, $\cup$, and $||$,
\begin{subequations} \label{energies}
\bea
\frac{\Omega_{\vee}}{\cal N} &=& \int_{u_c}^\infty du\, u^{3/2}\frac{u^5+b^2u^2-\frac{1}{2}[(bu_c+d)^2-n_4^2]}{\sqrt{g(u)}} 
+ \frac{n_4 u_c}{6}- \frac{\mu n}{2}   \, , \hspace{3cm}\\[1.5ex]
\frac{\Omega_{\cup}}{\cal N} &=& \int_{u_0}^\infty du\, u^{3/2} \frac{u^5+b^2u^2-\frac{1}{2}\frac{\eta^2}{\eta^2 +1}(u_0^5+b^2u_0^2)}
{\sqrt{u^8+b^2u^5-(u_0^8+b^2u_0^5)-(u^3-u_0^3)\frac{\eta^2}{\eta^2 +1}(u_0^5+b^2u_0^2)}} -\frac{\mu n}{2} \, ,\label{OmegaB}\\[1.5ex]
\frac{\Omega_{||}}{\cal N} &=& \int_0^\infty 
du\,u^{3/2} \frac{u^5+b^2u^2+\frac{1}{2}\frac{(3b\mu)^2}{\sinh^2 z_\infty}}{\sqrt{u^8+b^2u^5+u^3 \frac{(3b\mu)^2}{\sinh^2 z_\infty}}}
-\frac{\mu n}{2}  \, .\label{OmegaXs}
\eea
\end{subequations}
The first free energy is taken from sec.\ \ref{sec:free} and describes 
baryonic matter, i.e., the flavor branes are connected and have a cusp
in their profile at the location of the pointlike D4-branes $u=u_c$. The second describes mesonic matter which we have worked out in
ref.\ \cite{Preis:2010cq}. In this case,
the flavor branes are also connected but have a smooth profile everywhere, in particular at tip of the branes $u=u_0$. The quantities $u_0$ and 
$\eta$, appearing in the free energy, have to be determined from two coupled algebraic equations.  At the baryon onset, we have $u_c=u_0$ and 
$n_4=0$. Using the explicit form of the equations for $u_0$ and $\eta$ \cite{Preis:2010cq}, this yields 
$(bu_c+d)^2-n_4^2=\frac{\eta^2}{\eta^2+1}(u_0^5+b^2u_0^2)$, which shows that at the onset $\Omega_\vee  = \Omega_\cup$. The 
third free energy is also taken from ref.\ \cite{Preis:2010cq} and describes chirally restored matter, i.e., the case of disconnected 
flavor branes. Here we have one algebraic equation which has to be solved numerically for the quantity $z_\infty$. Since different solutions
constitute the state of lowest free energy in different regions of the phase diagram, there is a first-order phase transition within the 
symmetric phase which can be associated with a transition into the LLL. While the phase of higher Landau levels (``hLL'') corresponds
to a nontrivial solution, the ``LLL'' phase has the trivial solution $z_\infty=\infty$, 
which leads to a simple expression for the free energy, 
\be
\frac{\Omega_{||}^{\rm LLL}}{\cal N} = \int_0^\infty du\,\sqrt{u^5+b^2u^2}-\frac{\mu n}{2}  \, .
\ee
Obviously, this free energy, like all three expressions in eqs.\ (\ref{energies}), is divergent. After subtracting the vacuum 
free energy $\Omega_\cup(\mu=0)$ all expressions become finite. Equivalently, we can simply take the pairwise difference of free energies 
to find the ground state at any given $b$ and $\mu$. The resulting phase diagram of this
numerical calculation is shown in fig.\ \ref{figPD}. 

\begin{figure}[t]
\begin{center}
\hbox{\includegraphics[width=0.47\textwidth]{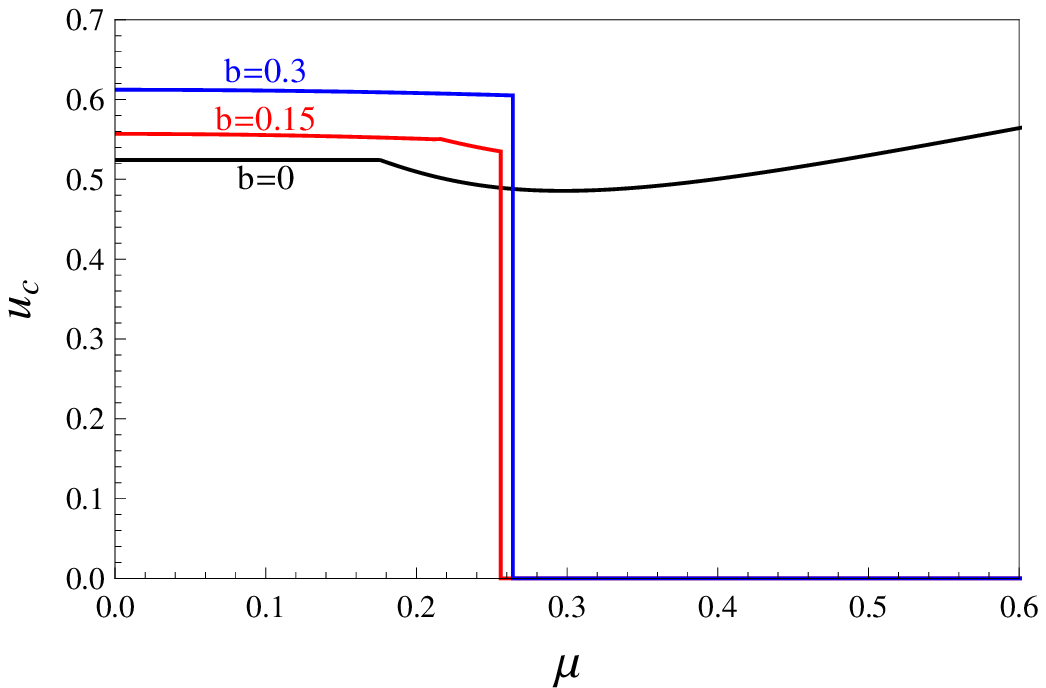}\hspace{0.3cm}
\includegraphics[width=0.47\textwidth]{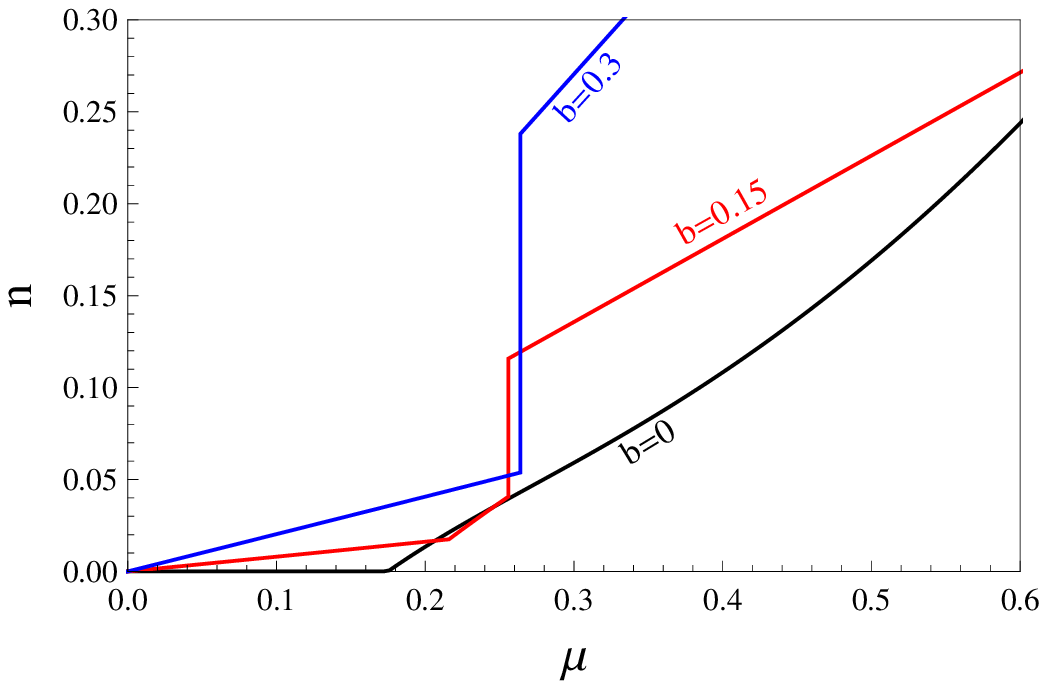}}
\caption{Left panel: location of the tip of the connected branes $u_c$ (denoted $u_0$ in the mesonic phase) 
as a function of $\mu$ for three values of the magnetic field, i.e., along three horizontal cuts through the phase diagram in the right panel
of fig.\ \ref{figPD}. One representative for each of the three qualitatively different cases is shown: baryon onset (small magnetic fields); 
baryon onset followed by chiral phase transition (intermediate magnetic fields); chiral phase transition (large magnetic fields). 
As fig.\ \ref{figPD} shows, for chemical potentials beyond the scale of the plot the system reenters the chirally broken phase, resulting in 
an additional first-order phase transition in all three cases. Right panel: baryon number density $n$ along the same cuts through the phase 
diagram. Through the axial anomaly the meson supercurrent produces a nonzero $n$ also in the mesonic phase for $b,\mu>0$.}
\label{figucn}
\end{center}
\end{figure}
For comparison, we have also included the phase diagram without baryonic matter from 
ref.\ \cite{Preis:2010cq}. In fig.\ \ref{figucn} we present the location $u_c$ of the tip of the connected branes and the 
baryon density along lines of constant magnetic field. The interpretation of the former deserves a comment. In the mesonic phase, $u_c$ is usually
identified with the constituent quark mass since it is the distance between the color and flavor branes. This is still true in the 
baryonic phase, but there it has a second meaning because $\frac{N_c u_c}{3}$ is the baryon mass. (Note that the factor $\frac{1}{3}$ 
originates from the geometry of the model and has nothing to do with the number of colors.)

\subsection{Observations}
\label{sec:discussion2}

The main observations can be summarized as follows. 
\begin{itemize}
\item The effect of baryonic matter at small magnetic fields is dramatic: it prevents the system from restoring chiral symmetry for any 
chemical potential. In doing so, it completely expels the ``hLL'' phase from the phase diagram, such that the only surviving 
chirally restored phase is the ``LLL'' phase.

\item The baryon onset line terminates at a critical endpoint $(\mu,b)\simeq (0.26,0.25)$
where it intersects the chiral phase transition line. As a consequence, at sufficiently 
large magnetic fields the mesonic phase is superseded by quark matter before baryonic matter can even be created. 

\item In the presence of baryonic matter, IMC plays an even more prominent role in the phase diagram:
at {\it any} given chemical potential $\mu \gtrsim 0.26$ a sufficiently large magnetic field induces chiral symmetry restoration.

\end{itemize}

Let us briefly comment on the first of these observations whose $b=0$ version was already made in ref.\ \cite{Bergman:2007wp}. 
The fact that chiral symmetry remains broken along the entire $\mu$ axis is an apparently puzzling result: in view of the similarity of the 
present non-antipodal version of the Sakai-Sugimoto model and
the NJL model, one might have expected chiral symmetry to be intact for sufficiently large chemical potentials. This was true {\it before} 
including baryonic matter, in which case the entire phase structure with magnetic field was in amazing agreement with 
NJL calculations \cite{Preis:2010cq}. However, in the presence of baryonic 
matter it is questionable to expect this agreement to persist for the following reason.  
In related NJL studies \cite{Frolov:2010wn,Inagaki:2003yi,Ebert:1999ht,Boomsma:2009yk} baryon number is a rather simple concept: 
the only degrees of freedom are quarks whose masses acquire a contribution from the chiral condensate. When the chemical potential is larger than 
this constituent quark mass, a nonzero quark number is generated. Then, baryon number is simply this quark number divided by $N_c$. This is not true 
in our present approach. Holographic baryons {\it are} different from a mere collection of $N_c$ quarks. This is clear from their construction 
and can easily be seen from the difference between the constituent quark mass in the mesonic phase $u_c$ and the baryon mass $\frac{N_cu_c}{3}$. 
Whether we should therefore call our quarks confined is debatable, but it is certainly a qualitative difference to the NJL model. 

To interpret the lack of chiral symmetry restoration, the following intriguing property of our baryonic matter at asymptotically large chemical 
potentials may be helpful. 
For simplicity, let us restrict ourselves to $b=0$. In this case, as demonstrated in appendix \ref{app2}, the free energy of the baryonic 
phase approaches 
the one of the chirally restored phase, 
\be
\frac{\Omega_{\vee}(b=0)}{\cal N} = -\frac{2}{7} \frac{\mu^{7/2}}{p^{5/2}} + {\cal O}(\mu^{5/2}) \, , \qquad 
\frac{\Omega_{||}(b=0)}{\cal N} = -\frac{2}{7} \frac{\mu^{7/2}}{p^{5/2}} \, , 
\ee
where $p = \Gamma\left(\frac{3}{10}\right)\Gamma\left(\frac{6}{5}\right)/\sqrt{\pi}$ (see fig.\ \ref{figpressure} in appendix 
\ref{app2} for a plot of the full numerical result).
Therefore, at asymptotically large $\mu$, baryonic matter and quark matter become thermodynamically indistinguishable. One may speculate that this 
is a consequence of the pointlike nature of our baryons: due to this property, our baryons can only overlap at infinite density. This suggests
that the expected transition to quark matter at finite $\mu$ is shifted to $\mu=\infty$ (which, curiously, is undone by a sufficiently large 
magnetic field). It would be interesting to see whether 
this is different for $N_f>1$, where baryons {\it can} be described by non-singular instantons.


\section{Summary and outlook}
\label{sec:summary}

We have discussed baryonic matter in a magnetic field, using the deconfined geometry of the Sakai-Sugimoto model where baryons are introduced
by D4-branes wrapped on the internal $S^4$. Our main focus has been the onset of baryonic matter and its effect on the 
chiral phase transition at zero temperature and finite chemical potential. 

The critical chemical potential for the onset of holographic baryonic matter increases monotonically with the magnetic field, saturating
at a finite value for asymptotically large magnetic fields. For subcritical chemical potentials the system is in the mesonic phase
with a meson supercurrent in the direction of the magnetic field. Because of the axial anomaly, the system has nonvanishing baryon number for all 
nonzero magnetic fields and chemical potentials. Due to the presence of the supercurrent the baryon onset is significantly ``delayed'' to 
larger chemical potentials. In contrast to real-world baryonic matter, the onset is a second-order phase transition.  Besides the holographic
study, we have computed the onset also in the Walecka model, a field-theoretical model for dense nuclear matter where the onset is a first-order
transition. Within this model, we have demonstrated that the onset depends strongly on the electric 
charge and the anomalous magnetic moment of the baryons. In most cases, however, the magnetic field favors baryonic 
matter because it decreases the baryon mass, although for neutral baryons it also decreases the (modulus of the) binding energy. In the holographic 
calculation there is no binding energy and the baryon mass is increased by the magnetic field. Our results indicate that this increase is closely 
related to the effect of MC. 

Without baryonic matter, chiral symmetry is restored for any given magnetic field at some sufficiently large chemical potential
\cite{Preis:2010cq}. With baryonic matter, this is only true for sufficiently large magnetic fields, where we have found that 
baryons play no role. 
For small magnetic fields there is a transition from mesonic to baryonic matter but no subsequent transition to quark matter. This enforces the 
unusual effect of IMC: although the magnetic field typically enhances a chiral condensate, it now {\it restores} chiral 
symmetry for any given chemical potential larger than, roughly speaking, the one where the baryon onset line and the chiral phase transition line
meet in a critical endpoint see fig.~\ref{figPD}.

As our comparison with the Walecka model has shown, one has to be very careful with drawing conclusions from the holographic results for 
the QCD phase structure and the interior of neutron stars, where the extreme conditions studied here might be realized. Most importantly, 
our holographic calculation has been restricted to the $N_c\to \infty$ limit, and there are indications that large-$N_c$ nuclear matter is very 
different from real nuclear matter. Therefore, generalizations to finite $N_c$ would be very interesting, as it has been done for instance 
in related D3/D7 models \cite{Bigazzi:2009bk}. On the other hand, in certain aspects our approach seems to be more realistic than 
widely used models of dense matter. 
In particular, it is interesting to compare our results to the NJL model. In our previous work \cite{Preis:2010cq} we have pointed out an amazing 
agreement with corresponding NJL phase diagrams \cite{Inagaki:2003yi,Boomsma:2009yk}. This agreement is lost after including baryonic matter. This is understandable 
since in the NJL model there are no baryons; baryon number is only generated by deconfined quarks. In the Sakai-Sugimoto 
model, however, baryons are clearly distinct from a set of $N_c$ deconfined quarks.

It would be interesting to extend our study to more than one flavor in order to describe more realistic baryonic matter.
Moreover, one might include superfluidity of nuclear matter, as suggested for the present model in ref.\ \cite{Rama:2011ny}. 
In the Sakai-Sugimoto model as well as in AdS/QCD approaches, it has also been suggested that -- even for vanishing magnetic field --
the ground state breaks rotational and/or translational invariance at sufficiently large baryon densities \cite{Domokos:2007kt,Chuang:2010ku,
Ooguri:2010xs,Bayona:2011ab}. 
It would be interesting 
to study the effect of such phases on our phase diagram.
We should also keep in mind that the quark matter phase considered here is not in a state expected from QCD. As indicated in 
sec.\ \ref{sec:context}, cold and sufficiently dense quark matter is a color superconductor in the CFL phase. This phase also breaks chiral 
symmetry at asymptotically large $\mu$, like our holographic baryons, but the mechanism is very different and heavily relies on the fact that 
$N_c=3$.

\begin{acknowledgments}
We thank L.\ Bonanno, P.\ Burikham, A.\ Gynther, K.\ Rajagopal, and M.\ Stephanov for valuable comments. 
This work has been supported by the Austrian science foundation FWF under project no.\ P22114-N16.
\end{acknowledgments}

\appendix 

\section{Solving the equations of motion}
\label{app1}

In this appendix we solve the equations of motion (\ref{EOM}). 
To some extent we can use the same method as in ref.\ \cite{Preis:2010cq} for the case without baryons. It is useful to rewrite the equations
as
\begin{subequations} \label{EOM3}
\bea
\frac{a_0'}{a_3'} &=& \frac{3ba_3+n_4}{3ba_0+d} \, , \label{EOM31} \\
\frac{a_0'}{u^3x_4'} &=& \frac{3ba_3+n_4}{k} \, , \label{EOM32} \\
\frac{a_3'}{u^3x_4'} &=& \frac{3ba_0+d}{k} \, . \label{EOM33}
\eea
\end{subequations}
We integrate eq.\ (\ref{EOM31}) to obtain
\be \label{a02}
\frac{3}{2}ba_0^2+da_0=\frac{3}{2}ba_3^2+n_4a_3+\kappa \, , 
\ee
where the integration constant $\kappa$ can be determined from evaluating this equation at $u=u_c$,
\be \label{kappa}
\kappa = \frac{u_c}{3}\left(d\pm\frac{bu_c}{2}\right)\, , 
\ee
where $a_3(u_c)=0$ and $a_0(u_c)=\pm \frac{u_c}{3}$ has been used. Using eqs.\ (\ref{EOM31}), (\ref{EOM32}), (\ref{a02}), and (\ref{kappa}), 
we compute 
\be 
a_3'^2-a_0'^2+u^3x_4'^2 = \frac{a_0'^2}{(3ba_3+n_4)^2}\left[(b u_c\pm d)^2 -n_4^2+\frac{k^2}{u^3}\right] \, .
\ee
Inserting this expression into the original equations of motion for $a_0$ and $a_3$ 
(obtained from taking the derivative with respect to $u$ on both 
sides of eqs.\ (\ref{EOMa0}) and (\ref{EOMa3})), one shows that these two equations are equivalent to 
\begin{subequations} \label{EOM4}
\bea
\partial_y a_0 &=& a_3(y)+\frac{n_4}{3b} \, , \\
\partial_y a_3 &=& a_0(y)+\frac{d}{3b} \, ,
\eea
\end{subequations}
where the new variable $y$ fulfils
\be
y' = \frac{3b u^{3/2}}{\sqrt{g(u)}} \, , 
\ee
with $g(u)$ defined in eq.\ (\ref{g}). Now, inserting $a_3'=y'\partial_y a_3 = y'\left(a_0+\frac{d}{3b}\right)$ into eq.\ (\ref{EOM33}) yields 
the solution for the embedding function,
\be \label{x4u}
x_4(u) = k\int_{u_c}^u\frac{dv}{v^{3/2}\sqrt{g(v)}} \, .
\ee
The system of coupled equations (\ref{EOM4}) for $a_0$ and $a_3$ can easily be solved. With $y(u_c)=0$ we have the boundary conditions 
$a_0(y=0)=\pm \frac{u_c}{3}$, $a_3(y=0)=0$, and the solutions become
\begin{subequations} \label{a0a3}
\bea
a_0(y) &=& \frac{1}{3b}\left[(d\pm bu_c)\cosh y + n_4\sinh y - d\right] \, , \label{a0y}\\[1.5ex]
a_3(y) &=& \frac{1}{3b}\left[(d\pm bu_c)\sinh y+n_4(\cosh y-1)\right]  \, . \label{a3y}
\eea
\end{subequations}
With $y_\infty\equiv y(u=\infty)$ the boundary conditions at the holographic boundary $a_0(y_\infty)=\mu$, $a_3(y_\infty)=\jmath$
can be used to determine $d$ and $n_4$,
\begin{subequations} \label{dn4}
\bea
d &=& -\frac{3}{2}b\mu - \frac{3}{2}b\left(\pm \frac{u_c}{3}-\jmath\,\frac{1+\cosh y_\infty}{\sinh y_\infty}\right) \, , \label{d1}\\[1.5ex]
n_4 &=& -\frac{3}{2}b\jmath + \frac{3}{2}b\left(\mu\mp \frac{u_c}{3}\right)\frac{1+\cosh y_\infty}{\sinh y_\infty} \, . \label{n41}
\eea
\end{subequations}
These expressions are valid for any $\jmath$. From the main text we know that the condition that $\jmath$ minimize the free energy is equivalent to 
$d=-\frac{3}{2}b\mu$. Inserting this value into eq.\ (\ref{d1}) yields the supercurrent given in the main text, eq.\ (\ref{j}), and, inserting the 
result for $\jmath$ into eq.\ (\ref{n41}), the density $n_4$ given in eq.\ (\ref{n4}). The results for $d$ and $n_4$ can then be inserted into 
eqs.\ (\ref{a0a3}) to obtain eqs.\ (\ref{solutions}).

Comparing eq.\ (\ref{n1}) with eq.\ (\ref{n41}) we obtain the total density 
\be
n = \frac{3}{2}b\left(\mu\mp \frac{u_c}{3}\right)\frac{1+\cosh y_\infty}{\sinh y_\infty} \, .
\ee
With the help of eqs.\ (\ref{dn4}) and $k$ from eq.\ (\ref{k21}) the solutions for $x_4$, $a_0$, $a_3$ in eqs.\ (\ref{x4u}) and (\ref{a0a3})
can be written in terms of $u_c$ and $y_\infty$ (and $\mu$ and $b$, which are externally fixed variables). 
The remaining two quantities, $u_c$ and $y_\infty$, need to be determined numerically from the following two coupled equations: 
firstly, the defining equation for $y_\infty$,
\be \label{yinf}
y_\infty = 3b\int_{u_c}^\infty \frac{ u^{3/2}du}{\sqrt{g(u)}} \, ,
\ee
and secondly, the equation that fixes the asymptotic separation of the flavor branes, $\ell/2 = x_4(\infty) - x_4(u_c)$, which reads
\be \label{ell}
\frac{\ell}{2} = k\int_{u_c}^\infty \frac{du}{u^{3/2}\sqrt{g(u)}} \, .
\ee

\section{Approximations close to the baryon onset}


\subsection{Vanishing B}
\label{app2a}

In this appendix we derive eq.\ (\ref{n4onset}), i.e., the behavior of the baryon density $n_4$ close to the onset at vanishing magnetic field.
For $b=0$ and with $y_\infty = 3by_\infty^0$, eqs.\ (\ref{yinfell}) can be written as
\begin{subequations}
\bea
(u_c^0)^{3/2}y_\infty^0 &=& \int_1^\infty\frac{u^{3/2}du}{\sqrt{u^8-1+\frac{n_4^2}{(u_c^0)^5}\left(u^3-\frac{8}{9}\right)}} \, , \label{beta}
\\[1.5ex]
\frac{(u_c^0)^{1/2}\ell}{2} &=& \sqrt{1+\frac{8}{9}\frac{n_4^2}{(u_c^0)^5}}\int_1^\infty
\frac{du}{u^{3/2}\sqrt{u^8-1+\frac{n_4^2}{(u_c^0)^5}\left(u^3-\frac{8}{9}\right)}} \, . \label{alpha}
\eea
\end{subequations} 
For chemical potentials close to, but above, the onset chemical potential $\mu^0_{\rm onset}=m_q^0$, the numerical results suggest the ansatz
\be
u_c^0 \simeq u_{c,{\rm onset}}^0 + \alpha \epsilon  \, , \qquad y_\infty^0 \simeq y_{\infty,{\rm onset}}^0 
+ \beta \epsilon \, ,
\ee
with $\alpha$, $\beta$ to be determined, and 
\be
\epsilon \equiv \mu-m_q^0 \, .
\ee
Inserting this ansatz into eq.\ (\ref{n4approx}) yields
\be \label{n4alpha}
n_4 \simeq \frac{3+\alpha}{3y_{\infty,{\rm onset}}^0}\,\epsilon \, .
\ee 
To compute $\alpha$ it is sufficient to consider eq.\ (\ref{alpha}), where $\beta$ does not appear. We are interested in the terms linear in 
$\epsilon$. The linear term on the left-hand side is easily obtained. On the right-hand side,
we subtract the constant term and neglect the quadratic term in the square root in front of the integral. If our ansatz is correct, the integral 
then must yield a linear term but, as written, also yields terms of higher order in $\epsilon$,
\be \label{eps}
-\frac{\alpha\ell}{4(u_{c,{\rm onset}}^0)^{1/2}}\,\epsilon \simeq \int_1^\infty
du\Big[\frac{1}{u^{3/2}\sqrt{u^8-1+v^2\epsilon^2\left(u^3-\frac{8}{9}\right)}}-\frac{1}{u^{3/2}\sqrt{u^8-1}}\Big] \, .
\ee
Here we have abbreviated 
\be
v\equiv \frac{3+\alpha}{3y_{\infty,{\rm onset}}^0(u_{c,{\rm onset}}^0)^{5/2}}  \, .
\ee
Let us for the following arguments denote the integral in eq.\ (\ref{eps}) by ${\cal I}$. One can check numerically that ${\cal I}$ 
is indeed linear in $\epsilon$ for small $\epsilon$. Obviously, we cannot proceed by naively expanding the integrand in $\epsilon$ since this 
procedure would miss the linear term. Instead, we employ the following trick. Neglecting higher order terms, ${\cal I}$ divided by $\epsilon$
should not depend on $\epsilon$ anymore, i.e., $\frac{\partial}{\partial\epsilon}\frac{\cal I}{\epsilon}\simeq 0$. We evaluate this equation by 
first rewriting the derivative with respect to $\epsilon$ as a derivative with respect to $u$,
\be\label{swapderivative}
\frac{\partial}{\partial\epsilon}\frac{1}{\sqrt{u^8-1+v^2\epsilon^2\left(u^3-\frac{8}{9}\right)}} = \frac{2\epsilon v^2\left(u^3-\frac{8}{9}\right)}
{u^2(8u^5+3v^2\epsilon^2)}\frac{\partial}{\partial u} \frac{1}{\sqrt{u^8-1+v^2\epsilon^2\left(u^3-\frac{8}{9}\right)}} \, . 
\ee
Then, with partial integration we obtain 
\be
\frac{\cal I}{\epsilon} \simeq -\frac{v}{12} -2\epsilon v^2 \int_1^\infty du\,\frac{1}{\sqrt{u^8-1+v^2\epsilon^2\left(u^3-\frac{8}{9}\right)}}
\frac{\partial}{\partial u}\frac{u^3-\frac{8}{9}}{u^{7/2}(8u^5+3v^2\epsilon^2)} \, ,
\ee
where the first term on the right-hand side is the boundary term. Now the left-hand side and the boundary term are 
constant in $\epsilon$ while the second term on the right-hand side is linear in $\epsilon$ and can thus be dropped (the integral is finite 
and yields a term constant in $\epsilon$, but no $\epsilon^{-1}$ term). Consequently, we have arrived at the very simple result 
${\cal I}=-\frac{v}{12}\epsilon$ which can be confirmed numerically and which we insert into eq.\ (\ref{eps}). The 
result is
\be \label{alpha1}
\alpha = \frac{3}{9y_{\infty,{\rm onset}}^0 (u_{c,{\rm onset}}^0)^2 \ell -1} \, .
\ee
With the same trick eq.\ (\ref{beta}) can be evaluated to obtain $\beta$. Here we are only interested in the baryon density for which we 
insert eq.\ (\ref{alpha1}) into eq.\ (\ref{n4alpha}). With $u_{c,{\rm onset}}^0$ and $y_{\infty,{\rm onset}}^0$ from eq.\ (\ref{uc0yinf0})
we obtain the final result given in eq.\ (\ref{n4onset}) in the main text. 

\subsection{Asymptotically large B}
\label{app2b}
Here we derive eq.\ (\ref{n4onsetbinfty}), i.e. 
the behavior of the usual baryon density $n_4$ close to the onset at asymptotically large magnetic fields. In this case, due to the supercurrent, the onset does not occur at the 
baryon mass but at twice the baryon mass, 
$\mu^\infty_{\rm onset}=2m_q^\infty$, where $m_q^\infty= u_{c,{\rm onset}}^\infty/3$ denotes the (dimensionless) constituent quark mass in a baryon at $b\to \infty$. 
As in the previous appendix, we first need to compute $y_\infty$ and $u_c$. With $y_\infty\to\infty$ for $b\to\infty$ we conclude from eq.\ (\ref{n4}) that
\be \label{n4inf}
n_4^\infty = \frac{3b}{2}\left(\mu-\frac{2u_c^\infty}{3}\right) \, , 
\ee
and thus we can write the second of eqs.\ (\ref{yinfell}) as
\be \label{separationbinfty}
\frac{\ell(u_c^\infty)^{1/2}}{2}=\sqrt{1-\left(\frac{n_4^\infty}{3bu_c^\infty}\right)^2}\int_1^\infty\frac{du}{u^{3/2}\sqrt{u^5-1+\left(\frac{n_4^\infty}{3bu_c^\infty}\right)^2}} \, .
\ee
This single equation can now be used to determine the behavior of $u_c^\infty$ at the onset. Again, as for $b=0$, the numerical result suggests the ansatz
\be \label{ucinf}
u_c^\infty \simeq u_{c,{\rm onset}}^\infty + \tilde{\alpha} \tilde{\epsilon} \, ,\qquad \tilde{\epsilon} \equiv \mu-2m_q^\infty 
\ee
with $\tilde{\alpha}$ to be determined. Inserting eq.\ (\ref{ucinf}) into eq.\ (\ref{n4inf}) yields
\be
n_4^\infty \simeq \frac{3b}{2}\tilde{v}\tilde{\epsilon} \, , \qquad 
\tilde{v}\equiv 1-\frac{2\tilde{\alpha}}{3}\, .
\ee
With the help of these expressions for $n_4^\infty$ and $u_c^\infty$ we see that to zeroth order in $\tilde{\epsilon}$ eq.\ (\ref{separationbinfty})
implies
\be
u_{c,{\rm onset}}^\infty=\frac{16\pi}{\ell^2} \left[\frac{\Gamma\left(\frac{3}{5}\right)}{\Gamma\left(\frac{1}{10}\right)}\right]^2 \,, 
\ee
from which the onset chemical potential in eq.\ (\ref{asymp}) is obtained. It is obvious that the left-hand side of eq.\ (\ref{separationbinfty}) has a linear term in $\tilde{\epsilon}$.
For the right-hand side, however, we need to apply a similar trick as in the previous appendix to extract the linear term. We write the linear terms of 
eq.\ (\ref{separationbinfty}) as
\bea\label{separationapprox}
\frac{\ell\tilde{\alpha}\tilde{\epsilon}}{4(u_{c,{\rm onset}}^\infty)^2} = \left[\frac{\partial}{\partial\tilde{\epsilon}}\int_1^\infty\frac{du}{\sqrt{u^5-1+
\frac{\tilde{v}^2\tilde{\epsilon}^2}{4(u_{c,{\rm onset}}^\infty)^2}}}\right]_{\tilde{\epsilon}=0}\tilde{\epsilon} \, .
\eea
Now, analogously to eq.\ (\ref{swapderivative}) we rewrite the differentiation with respect to $\tilde{\epsilon}$ by a differentiation with respect to $u$ and
compute the integral via partial integration. Then, for $\tilde{\epsilon} = 0$ only the boundary term survives, and we can easily compute the result for $\tilde{\alpha}$,
\be
\tilde{\alpha} = \frac{12}{8-15(u_{c,{\rm onset}}^\infty)^{1/2}\ell} 
\ee
Inserting this into eq.\ (\ref{ucinf}) and the result into eq.\ (\ref{n4inf}) yields $n_4^\infty$ as given in eq.\ (\ref{n4onsetbinfty}) in the main text.

\section{Asymptotics at large chemical potential}
\label{app2}

\begin{figure} [t]
\begin{center}
\includegraphics[width=0.45\textwidth]{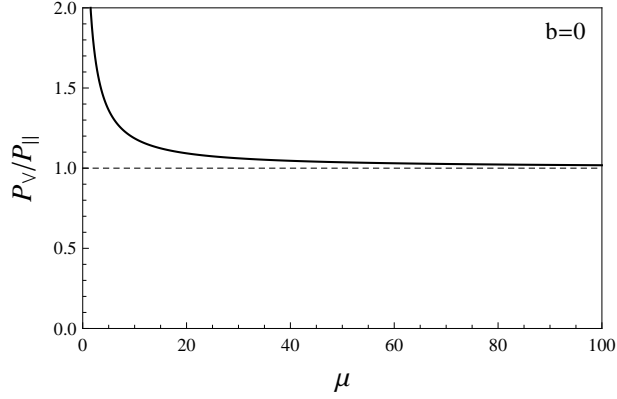}
\caption{Ratio of the pressures of baryonic matter $P_\vee$ and quark matter $P_{||}$ at vanishing magnetic field as a function of $\mu$. 
The plot shows that baryonic matter always has larger pressure, i.e., chiral symmetry is never restored, but at asymptotically large $\mu$
baryonic and quark matter become indistinguishable. (Note the huge $\mu$ scale compared to the phase diagrams in fig.\ \ref{figPD}.)}
\label{figpressure}
\end{center}
\end{figure}

In this appendix we compute the behavior of the baryonic phase at asymptotically large $\mu$ and $b=0$.
The numerical evaluation shows that $y_\infty \propto\mu^{-3/2}$, $u_c\propto \mu^0$ for $\mu\to\infty$. Hence we make the ansatz
\be
y_\infty \simeq \frac{3b\,p^{5/2}}{\mu^{3/2}} \, , 
\ee
where $p$ is a number which we shall now determine. We have extracted the linear behavior in $b$ which is necessary to take the $b\to 0$ limit. 
Inserting this ansatz into eq.\ (\ref{n4}) yields
\be
n_4 \simeq \left(\frac{\mu}{p}\right)^{5/2} \, , 
\ee
and thus the equation for $y_\infty$ (\ref{yinfell}) becomes 
\be
p^{5/2} \simeq \mu^{3/2} \int_{u_c}^\infty\frac{u^{3/2}du}{\sqrt{u^8 +\frac{\mu^5}{p^5}\left(u^3-\frac{8}{9}u_c^3\right)}}
=\frac{1}{u_c^{3/2}}\int_{1/\mu}^\infty\frac{x^{3/2}dx}{\sqrt{x^8+\frac{1}{p^5u_c^5}\left(x^3-\frac{8}{9\mu^3}\right)}} \, , 
\ee
with the new integration variable $x=\frac{u}{u_c\mu}$. Now we can approximate the lower boundary by 0 and neglect the term $\propto \mu^{-3}$ in the
denominator of the integrand. Then, performing the resulting integral, we see that $u_c$ drops out of the equation and we obtain 
\be
p = \frac{\Gamma\left(\frac{3}{10}\right)\Gamma\left(\frac{6}{5}\right)}{\sqrt{\pi}} \, .
\ee
Next we can compute the free energy. From the general expression (\ref{Orewrite}) we find with the above asymptotic relations,  the same
change of integration variable, and recalling that $n=n_4$ at $b=0$,
\be
\frac{\Omega_{\vee}}{\cal N} \simeq \mu^{7/2} u_c^{7/2}\int_0^\infty dx\sqrt{x^5+\frac{1}{p^5u_c^5}} -\frac{\mu^{7/2}}{p^{5/2}} 
=\frac{2}{7}(\Lambda\mu u_c)^{7/2}-\frac{2}{7}\frac{\mu^{7/2}}{p^{5/2}} = \frac{\Omega_{||}^{\rm hLL}}{\cal N} \, .
\ee
Here we have performed the integral explicitly with a cutoff $\Lambda$ at the upper boundary and noticed that the asymptotic $b=0$ result of the 
baryonic phase is exactly the same as the full $b=0$ result for the ``hLL'' phase with restored chiral symmetry, see appendix D of 
ref.\ \cite{Preis:2010cq}.  At non-asymptotic values of $\mu$, the free energies of the two phases differ. In fig.\ \ref{figpressure}
we plot the ratio of the two corresponding pressures $P=-\Omega$, obtained from the numerical result.

\bibliographystyle{ropipt}
\bibliography{refs1}

\end{document}